**Opting Out of Generative AI: a Behavioral Experiment on the Role of Education in Perplexity AI Avoidance**

Ulloa, Roberto[1]; Kulshrestha, Juhi[2]; Kacperski, Celina[1,3]*

[1] Cluster of Excellence "The Politics of Inequality", University of Konstanz, Konstanz, Germany
[2] Department of Computer Science, Aalto University, Espoo, Finland
[3] Cluster of Excellence "Collective Behaviour", University of Konstanz, Konstanz, Germany

*Corresponding author: celina.kacperski@uni-konstanz.de

**Statements and Declarations**

**Acknowledgments**
We thank the Seek2Judge team for their support.

**Funding**
Deutsche Forschungsgemeinschaft (DFG – German Research Foundation) under Germany's Excellence Strategy – EXC-2035/1 – 390681379

**Competing interests**
The authors declare that they have no competing interests.

**Abstract**

The rise of conversational AI (CAI), powered by large language models, is transforming how individuals access and interact with digital information. However, these tools may inadvertently amplify existing digital inequalities. This study investigates whether differences in formal education are associated with CAI avoidance, leveraging behavioral data from an online experiment (N = 1,636). Participants were randomly assigned to a control or an information-seeking task, either a traditional online search or a CAI (Perplexity AI). Task avoidance (operationalized as survey abandonment or providing unrelated responses during task assignment) was significantly higher in the CAI group (51%) compared to the search (30.9%) and control (16.8%) groups, with the highest CAI avoidance among participants with lower education levels (~74.4%). Structural equation modeling based on the theoretical framework UTAUT2 and LASSO regressions reveal that education is strongly associated with CAI avoidance, even after accounting for various cognitive and affective predictors of technology adoption. These findings underscore education's central role in shaping AI adoption and the role of self-selection biases in AI-related research, stressing the need for inclusive design to ensure equitable access to emerging technologies.

*Keywords*:  digital divide, conversational AI, UTAUT, AI adoption, online behavior, online experiments

**Introduction**

The rapid advancement of generative artificial intelligence (genAI) has transformed the way individuals interact with digital information (Stokel-Walker & Van Noorden, 2023). Some of these tools are designed to provide users with natural and dynamic means of searching for and retrieving information in a natural language dialogue (Radlinski & Craswell, 2017). Unlike traditional search engines, which return information on specific keywords in "single shot" interfaces, conversational search AIs such as Perplexity AI or Bing Copilot[1] ask follow-up questions, clarify intent, and personalize responses based on previous prompts and answers. Similarly to traditional search engines, they can provide sources for their information. Since large language models (LLMs) captured the public interest with the advent of ChatGPT, conversational AI (CAI) has been integrated into applications for, among others, customer service, e-commerce, and education, enhancing user experience by delivering more tailored and dynamic responses to our information needs (Casheekar et al., 2024). Studies have found that users often prefer conversational search systems, possibly due to improved speed of task completion, efficiency, and user-friendliness (Dubiel et al., 2018; Kaushik & Jones, 2023).

Despite this apparent preference for more dynamic information-seeking tools, questions about their accessibility and equitable use have recently emerged (Kacperski et al., 2025; Khowaja et al., 2024). The digital divide concept is central to understanding the challenges faced by users of emerging technologies like CAI; specifically, the concept refers to the disparities in access to, use of, and benefit from digital technologies among different social groups (Lythreatis et al., 2022; Nguyen et al., 2021). Research consistently shows that older adults and women, as well as individuals with lower socio-economic backgrounds and education, tend to report and display lower levels of technology acceptance (Cruz-Jesus et al., 2016; Elena-Bucea et al., 2021; Kacperski et al., 2025). Differences in the willingness to use CAI could exacerbate existing skill gaps and lead to disadvantaged individuals missing out on opportunities for productivity improvement and personalized learning (Capraro et al., 2024).

---

[1] The most known conversational AIs only introduced search capabilities much lager, e.g., ChatGPT in October 2024 and Gemini in March 2025.

Our study focuses on how individuals' educational attainment influences the behavior of engaging with or avoiding a task to use a fully fledged CAI, compared to two other tasks: searching online in the traditional way and a control condition. We focus our analysis on education primarily due to its policy relevance and malleability. Unlike age or gender, which are immutable characteristics, educational disparities can be meaningfully addressed through interventions—such as targeted digital literacy programs, simplified user interface design, or tailored onboarding experiences. Moreover, the role of education in technology adoption remains underexplored within the UTAUT2 framework (for a review, see Alsharhan et al., 2024). While age and gender are routinely included as moderators, educational attainment is often omitted, partly due to cross-cultural variation and the difficulty of accounting for informal or lifelong learning.

Finally, recognizing that participation in surveys about emerging technologies typically skews toward individuals with higher educational attainment, who see themselves as beneficiaries of digitalization (Grewenig et al., 2023) and that intrinsic motivation is a major driver to participation in research studies (Silber et al., 2023), it is essential to examine how self-selection biases might influence the research trajectory of CAI development. If research predominantly represents the perspectives of digitally empowered citizens, CAI could evolve in a way that further excludes lower-educated individuals by prioritizing the needs and experiences of those already at an advantage, underscoring the urgent need for inclusive research designs and targeted interventions to ensure equitable CAI adoption and use.

**The role of education in AI and CAI adoption**

In recent years, there has been a rapid increase in research examining individuals' willingness to adopt artificial intelligence (AI) across diverse fields, including health care, computational tasks, and scientific research (for reviews, see Kabalisa & Altmann, 2021; Kelly et al., 2023; Radhakrishnan & Chattopadhyay, 2020). More recently, attention has turned to understanding the adoption of conversational AI tools powered by newer-generation genAI models (see, e.g., a review on privacy, security, and trust in CAI; Leschanowsky et al., 2024). In general, there appears to be a consensus that the adoption of AI is, among other things, dependent on a variety of demographic and socio-economic factors: age, gender, income, education, and political stance (e.g., Bentley et al., 2024; Eder & Sjøvaag, 2024; Kacperski et al., 2025; C. Wang et al., 2024). Still, a review of studies on chatbot adoption reported that moderators, especially demographics, were not considered in 63% of reviewed studies, with age, gender, and technology experience being the most employed (Alsharhan et al., 2024).

Studies do indicate that advanced educational degrees have been associated with higher AI knowledge and skills (Eder & Sjøvaag, 2024; C. Wang et al., 2024), reliance on AI recommendations (Biswas & Murray, 2024), enthusiasm towards and support for developing AI (Zhang & Dafoe, 2019), intention to adopt due to higher AI self-efficacy (Hong, 2022), trust and expectations about usefulness (Araujo et al., 2020), and adoption of transcription and translation tools (Goldenthal et al., 2021), while those with lower education are more negative, or, at best, more indifferent or ambiguous (Bao et al., 2022). Specifically with regards to CAI, the Pew Recent Center reported that US adults with higher educational degrees are more likely to report having used ChatGPT (McClain, 2024). In a purely academic context, more years of education positively affected reported usage and attitudes towards CAI (Stöhr et al., 2024). In the tourism sector, individuals with bachelor's and above degrees reported higher acceptance of hotel-assistant CAIs, possibly due to a better understanding of their capabilities (Tavitiyaman et al., 2022).

However, the above adoption studies draw conclusions from self-reported survey data. Actual behavioral measures are so far lacking: to the best of our knowledge, only two studies have previously explored the association of socio-demographic factors and CAI adoption with behavioral data, both finding education to be associated with initial adoption: one study tracked German users' ChatGPT usage behavior with digital traces (Kacperski et al., 2025); another study used Bing search queries related to ChatGPT as a proxy for its access and usage across the US (Daepp & Counts, 2024).

**UTAUT2 and CAI adoption**

To better understand the interplay of various acceptance determinants and education, we draw on the Unified Theory of Acceptance and Use of Technology (UTAUT, Venkatesh et al., 2003, updated in 2012 to the UTAUT-2), a widely used framework for studying technology adoption. The UTAUT2 posits that constructs such as performance expectancy, effort expectancy, social influence, facilitating conditions, hedonic motivation, and habit play key roles in shaping behavioral intention and that behavioral intention (BI) predicts actual behavior, a premise central to many technology adoption studies. The underlying rationale is that individuals' intentions, which encompass their motivation and readiness to engage with a particular technology, are the most direct and observable precursors to action. Recently, a research agenda employing the UTAUT2 to study AI adoption has been proposed (Venkatesh, 2022), and the UTAUT2 model has started to be well-employed for this purpose (e.g., see reviews of Kelly et al., 2023; Leschanowsky et al., 2024; Radhakrishnan & Chattopadhyay, 2020). For example, Leschanowsky et al. (2024) found that six studies used the framework in a review related to CAIs' privacy, security, and trust. Additionally, studies have found evidence of the relationship of UTAUT2 variables such as performance and effort expectancy with CAI adoption specifically (Budhathoki et al., 2024; Ma & Huo, 2023; Vimalkumar et al., 2021).

      The UTAUT2 considers moderators in line with literature on the digital divide: it posits that specific subgroups are less likely to engage with novel technologies, particularly those who are already at a disadvantage, such as older adults, women, or individuals with lower education (Venkatesh et al., 2016). Evidence in the UTAUT2 literature supports this position, demonstrating that education shapes how participants perceive the usefulness and ease of use of technologies like internet use (Niehaves & Plattfaut, 2010). Education has been shown to be a significant moderator in the adoption-perception-use relationship towards Facebook when Facebook first emerged, even as all other UTAUT2 factors were significant (Liew et al., 2014). The adoption of internet banking and mobile internet banking app usage, for example, has also been found to be more commonly undertaken by users with higher education due to higher performance expectancy, trust, and locus of control (Abu-Shanab, 2011). At the same time, some partial or null effects are also reported in the literature for the moderating effect of education on UTAUT2 variables, for example, for mobile app adoption and mobile phone data usage, where at most, it moderated social influence (Hew et al., 2015; Li et al., 2014). In a study of the use of government internet services, education had a significant main effect, but when UTAUT2 factors, especially performance expectancy, and experience, were included, the relationship

disappeared (van Dijk et al., 2008). It suggests that education's moderating role in technology adoption is context-dependent and not as applicable to certain specific technological innovations.

Aside from the UTAUT2 items, researchers have recently developed and tested other scales to capture more specific attitudes and affective responses toward CAIs. Perrig et al. (2023) assessed the psychometric properties of trust scales in AI contexts, finding that trust and distrust are best treated as separate constructs and recommending refined versions of existing measures. Wang & Wang (2022) developed and validated a multi-item scale for AI anxiety, showing that it reliably predicts motivated learning behavior and captures general unease about AI's societal implications. Said et al. (2023) demonstrated that individuals' knowledge and confidence in their understanding of AI shape their perceptions of its risks and benefits, with more knowledgeable users showing a tendency toward risk blindness and more confident users emphasizing benefits.

**Research agenda**

We conducted an online experimental study in which participants were randomized into a control group, a group tasked with performing a traditional online search, and a group instructed to use a CAI (Perplexity AI) to inform themselves about a policy. A disproportionately high attrition rate occurred in the CAI group - 377 out of 736 (~51%) of the assigned participants - and the abandonment of the survey at the moment of the task assignment is an indicator of avoidance of the task. We thus formulated a first explorative hypothesis to support our observation statistically:

**H1:** Participants are significantly more likely to disengage at the moment of task assignment if assigned to use a CAI compared to control and traditional online search tasks.

To better understand this behavior, we turned to prior research, which has found in behavioral data that usage of CAI is significantly associated with user socio-demographics (Kacperski et al., 2025) and might be fueling an AI digital divide concerning educational levels (Daepp & Counts, 2024). We thus tested differences in education on CAI avoidance:

**H2:** Participants with lower education levels are significantly more likely to disengage at the moment of task assignment if allocated to the CAI group.

To explore the mechanisms driving the avoidance in the CAI group, we leveraged data collected in the same sample before group assignment: we asked UTAUT2 variables (Venkatesh et al., 2012) and variables related to CAI affect such as anxiety, optimism, and pessimism. We focus on associations with educational levels.

**H3:** Higher education levels are positively associated with UTAUT2 constructs and positive affective variables but negatively associated with negative affective variables (i.e., anxiety, pessimism, disadvantage).

**H4:** UTAUT2 constructs and positive affective variables are negatively associated with avoidance, and negative affective variables will be positively associated with CAI avoidance.

We then conducted an exploratory analysis using structural equation modeling to fit the UTAUT2 (Unified Theory of Acceptance and Use of Technology) framework. Our goal was to examine if education remains relevant as a predictor of CAI avoidance after introducing it into the framework or whether factors included in the UTAUT2 will explain away all avoidance effects.

**H5:** Participants with lower educational levels will have higher CAI avoidance even after controlling for all other variables included in the UTAUT2 framework.

We further assessed the robustness of education as a predictor of CAI avoidance by conducting a LASSO regression (using 10-fold cross-validation) to select relevant variables among demographics and affective scales (Perrig et al., 2023; Said et al., 2023; Y.-Y. Wang & Wang, 2022).

**H6:** The LASSO procedure will consistently select educational attainment as a key predictor of CAI avoidance. Affective and demographic variables are included in the model.

We contribute to the literature in two ways: firstly, we focus on the under-researched variable of educational level and its effect on the avoidance of a specific AI technology, CAI; this gives insights into the existence of a digital divide (Alsharhan et al., 2024). Secondly, unlike most studies that rely on self-reports of adoption or report behavioral intention only (for a review, see Kelly et al., 2023), we report a behavioral metric of CAI engagement when faced with the task of using Perplexity AI to search for information, reducing errors stemming from social desirability and false recall, and improving predictive validity.

## Methodology

**Experimental design**

We used data collected in the Seek2Judge project, which was reviewed by the institutional ethics board of the University of Konstanz. Participants were part of a larger study on online search and political attitude change and were sampled from an online market research panel by European panel provider Bilendi GmbH. They received a letter of information, gave explicit consent, and then completed an initial survey in August 2023. The initial survey collected data on socio-demographics and items related to political interests. Individuals who completed the initial survey were invited and consented to participate in the experiment conducted over one week, from September 5th to September 13th, 2023.

Before answering the items, participants were introduced to the topic: "*Artificial intelligence (AI) is increasingly used in conversational systems like ChatGPT, BingChat, Bard to enable human-like conversations, often in online chats. The following questions will focus on this application of AI, which we will always call "Chatbot" in the questions.*" An image of the ChatGPT interface was provided to illustrate a CAI. Later questions also included another reminder, "*In the following questions the term "Chatbot" refers to AI-supported conversational systems such as ChatGPT, BingChat or Bard*."

Participants then completed the baseline survey, which included adapted UTAUT2 items (Venkatesh et al., 2012) and a set of affective CAI-related items introduced in recent literature (Perrig et al., 2023; Said et al., 2023; Y.-Y. Wang & Wang, 2022); see Appendix G. They were then randomly assigned to one of three groups, namely (1) control, (2) search, and (3) CAI. The control group was asked to think about a German renewable energy transition policy, the *Erneuerbare Energien Gesetz* (EEG), consider its pros and cons and then write their opinion. They were asked to spend 5 minutes to complete this task, during which they could not progress in the survey. The search group was instructed to search online to inform themselves about the EEG and the pros and cons of its introduction and to provide resources and links they found online; they could not proceed until 5 minutes had elapsed. The CAI group was directed to interact with a CAI (Perplexity AI) about the EEG and the pros and cons of its introduction through a provided link that opened a new tab with the Perplexity interface. Here, while initially, the 5-minute was mandatory, the timer and the restriction were removed as attrition was disproportionately high in this group[2]. We find no evidence that the removal of the timer had an effect; indeed, CAI avoidance increased after the timer was removed across the tested models—likely due to motivational effects (e.g., initial participants tend to be more motivated and less likely to abandon the survey). No interactions with the groups were found (Appendix C1). In all models, we add a control variable for the presence of the timer (binary). The CAI group had to provide the URL of the conversation log. After completing the task, participants filled out further survey items about the EEG.

---

[2] We monitored the functioning of perplexity.ai and did not observe any changes in their operation.

The chosen CAI, Perplexity AI, is powered by a large language model (OpenAI's GPT-3.5 by the time of data collection; the same underlying model used for ChatGPT at the time) and generates responses based on search results. It was also launched around the same time as ChatGPT, in December 2022. It was chosen because, at the time of data collection, (1) it did not require a log-in account to initiate a conversation, (2) it was one of the few publicly available CAIs with incorporated backend web-search to provide current sources, which we deemed important for trust and (3) it generated a shareable link to the conversation log, facilitating data collection.

**Participants**

Our final sample consisted of N=1636 German-speaking residents of Germany between 25 and 64 years old: 458 in the control group, 444 in the search, and 734 in the CAI. Of the final sample, 780 (47.7%) identified as women and 856 (52.3%) as men (the German population skews 50.3% female), so our sample includes more men than the general population. The participants' median age was 48 (mean=46.8, SD=11.95); 20.1% between 25 to 34, 24% between 35 to 44, 21.8% between 45 to 54, and 34.1% between 55 to 64. The figures for the German population for the same age cohorts are 24.7%, 25.1%, 23%, and 27.2% – suggesting that our sample might be somewhat older than the general population. Educational attainment was measured using the education scale from the European Social Survey and then summarized into four categories: low (elementary school or equivalent, n=192, 11.7%), medium (low secondary

Opting Out of Generative AI                                                                 13

education, n=509, 31.1%), high (high school/vocational certificate or equivalent, n=427, 26.1%), and very high (university degree, n=508, 31.1%). The German population has more individuals in the low (18.9%) and high (35.1%) groups and fewer in the middle (25.7%) and very high (20.3%) groups. For distribution across intervention groups, see Appendix A. The results presented in this study are robust across both weighted and unweighted analyses (see below). German population statistics are sourced from the Federal Statistical Office (Statistisches Bundesamt, 2024).

Following Austin (2009), an absolute standardized mean difference (SMD) under 0.1 typically suggests adequate covariate balance across treatment groups. Although all absolute SMDs in our study are below this benchmark, values for women (0.094) and those with high education (0.094) are close to it (Appendix B). To address these discrepancies and deviations from the German population distribution, we conducted analyses using two separate weighting strategies, applied independently to each intervention group: poststratification weights (based on age, gender, and education) and raked (and trimmed) weights (derived from the marginal means of the same variables). The poststratification weighting produced three extreme weights (over 5 times the median) and a coefficient of variation (CV) of 0.59. Raked-trimmed weighting yielded no extreme weights, a CV of 0.50, and trimmed only 1.04% of weights; for the CAI group, the CV is .43. High CVs (such as those above >.5) are associated with loss of efficiency (reduced effective sample size) inducing larger standard errors and variance, i.e., tests are more conservative and more care should be exerted when interpreting non-significant results. Further details concerning weight calculation (and diagnostics) can be found in Appendix B. We report the raked and trimmed weighted results in the main text; unweighted and poststratification weighted analyses, showing negligible differences, are included in Appendix C, D, and F.

**Measurements**

Our dependent variable, task avoidance, is a binary behavioral measure (1 = avoidance; 0 = engagement). Specifically, task avoidance was coded as 1 if participants met any of the following conditions: (a) in the control task, they submitted unrelated text rather than providing an opinion about EEG (or abandoned the survey during the task interval, n=8); (b) in the search task, they either failed to provide any links or submitted unrelated content as a resource (or abandoned the survey during the task interval, n=60); and (c) in the CAI task, they did not submit a valid link to Perplexity AI; effectively, all had to abandon the survey because the provision of the link was a compulsory field. Conversely, participants who fulfilled the task requirements were coded as 0 (i.e., engaged).

Participants responded to UTAUT2 items assessing behavioral intention (BI), performance expectancy, effort expectancy, social influence, facilitating conditions, hedonic motivation, and habit using a 7-point Likert scale ranging from "strongly disagree" to "strongly agree." They also reported the different CAI they had used in the past (referred to as CAI diversity; only 37 participants reported having used perplexity AI previously) and the types of usage (referred to as applications, e.g., writing or programming). For the UTAUT2 framework, we derived a binary experience variable from the reported applications, indicating whether participants had any prior usage. Finally, we included affective scales related to CAI trust (n=2. $\alpha$ = .91; Perrig et al., 2023) and anxiety (n=3, $\alpha$ = .71; Wang & and Wang, 2022)—which we averaged to have a single indicator, as well as single-items of negativity, optimism, fear, perceived advantages, and perceived disadvantages (Said et al., 2023). A complete list of items is available in Appendix G.

**Statistical models**

To examine the treatment effects on avoidance, we conducted a logistic regression with task avoidance as the dependent variable. The analysis compared group assignments (control, search, CAI) moderated by education while controlling for age and gender, attentiveness (measured via three attention checks), and the timing condition (whether the survey was completed with the timer in place or after its removal). Age and attentiveness were scaled and centered.

We used the R lavaan package (Rosseel, 2012) to fit structural equation models (SEMs) to examine CAI avoidance (DV) using the UTAUT2 framework. We modeled behavioral intention (BI) influenced by UTAUT2 variables (performance expectancy, effort expectancy, social influence, facilitating conditions, hedonistic motivation, and habit), demographics (gender, age, and education), and experience. Following Venkatesh et al. (2012), moderation analysis included simple interaction between demographics and experience with UTAUT2 constructs: facilitating conditions, hedonistic motivation, and habit. In turn, BI, demographics, experience and facilitating conditions, hedonistic motivation, and habit served as predictors of CAI avoidance. Attentiveness was included as a control variable in the BI and avoidance modeling, while the timing condition was only included in the avoidance modeling. All predictors are scaled and centered for easy comparison. Figure 3 illustrates our model.

We use the WLSMV (Weighted Least Squares Mean and Variance adjusted) estimator, as recommended by Beauducel & Herzberg (2006), for models with categorical or ordinal outcomes to fit the SEMs. Binary outcomes can be interpreted as ordinal (engagement < avoidance), and WLSMV handles it accordingly. Compared to logistic regression (not supported in lavaan), WLSMV estimates tend to be more conservative, as the method penalizes extreme values and adjusts for distributional assumptions. However, the WLSMV estimates are not directly comparable to those from logistic regression (which are in log-odds), as they represent changes in the latent thresholds that separate the binary categories (avoidance vs. engagement), generally resulting in smaller coefficient values.

Social influence, habit, and behavioral intention were each measured using the single survey item with the highest factor loading, as reported in the UTAUT2 framework (Venkatesh et al., 2012). In our main structural equation modeling (SEM) analyses, we treat the reported score on the single item (manifest indicator) as the observed measure of the construct. We also conducted analyses using fixed factor loadings from the original UTAUT2 as a robustness check, which yielded highly consistent results (Appendix D4).

In all SEM models, we standardize all latent variables to unit variance and ensure that the factor loadings for latent variables are constrained to be equal across groups, i.e., the relationship between observed UTAUT2 indicators (survey items) and their respective latent constructs remains consistent across all experimental groups (control, search, CAI). Although we only report on the CAI group, this approach helps improve the precision and reliability of the latent constructs by leveraging data from all groups during the model estimation.

## Results

Avoidance rates for the control, search, and CAI are 16.8% (N=77), 30.9% (N=137), and 51% (N=374), respectively; statistically, we find support for **H1**: CAI avoidance was higher than search avoidance, OR = 2.29, 95% CI [1.75, 2.99], $p < .001$, and search avoidance was higher than avoidance in the control group, OR = 2.25, 95% CI [1.59, 3.19], $p < .001$ (Appendix C1).

We find support for the hypothesis that participants with lower education levels are more likely to avoid the CAI task (**H2**), as indicated by a significant interaction between linearized education and CAI group assignment; OR = 0.34, 95% CI [0.17, 0.68], *p* = 0.002 (Appendix C2). Additionally, we find a significant interaction between linearized education and search group assignment; OR = 0.36, 95% CI [0.16, 0.80], p = 0.012. Figure 1 illustrates these interactions; the observed interaction between the quadratic education term and search group assignment is statistically significant (p = 0.009; Appendix C3). Exploratively, we included age and education as moderators (Appendix C4), finding significant interactions between age and the CAI group (*p* = .001), age and the search group (*p* < .001), gender and the CAI group (*p* = .019), but not gender and the search group. In this model, the interaction between education and the search group remain significant (*p* = .025), while the interaction between education and the CAI group is significant for the unweighted model (*p* = .006) and the poststratification model (*p* = .024) but not for the trimmed and raked model (*p* = .056).

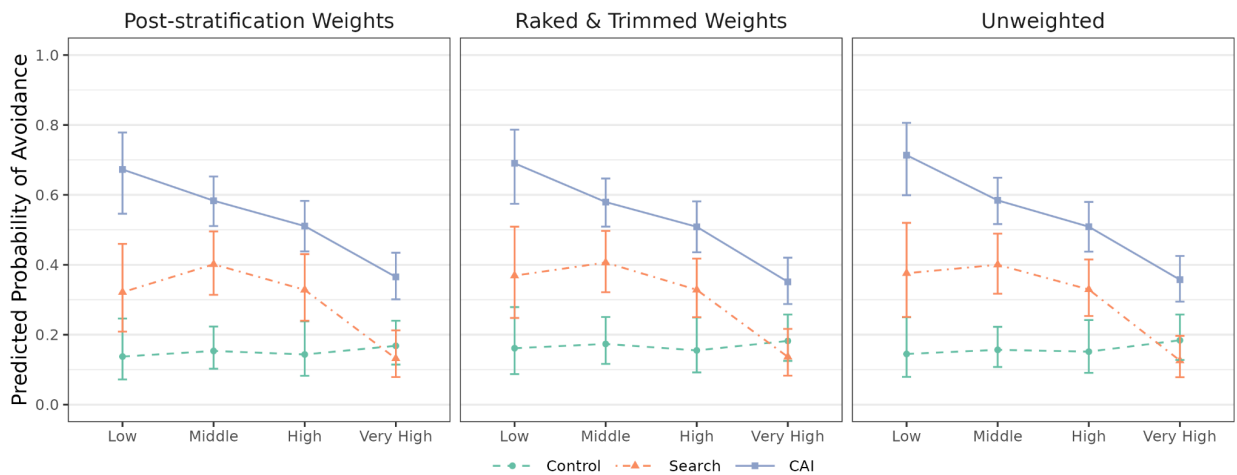

**Figure 1. Predicted avoidance probability by education level and experimental group under different weighting schemes.** The left panel shows unweighted estimates, the middle panel applies raking with trimming, and the right panel uses post-stratification weights. The estimates are calculated using marginal means via the emmeans R package (Lenth, 2024). Error bars represent 95% confidence intervals.

Table 1 presents the distributions of study variables across educational groups. We find support for **H3**: Higher education is positively associated with higher scores on all UTAUT2 variables (all $p_{adj}$ < .001)—including all metrics for experience—and with more positive affective responses toward CAIs (trust, optimism, and expectations of profiting from AI), (all $p_{adj}$ < .001), while it is negatively associated with lower scores on only one negative affective scales (i.e., pessimism, $p_{adj}$ (X²) =.008)—other negative scales (anxiety, negativity and perceived disadvantages from AI) were not significant (i.e., neither support nor contradict **H3**).

**Table 1. Variable distribution across education.** Each of the three sections in the table corresponds to variables associated with experience, UTAUT2, affective, and text metrics. $p_{adj}$ (X²) denotes the Bonferroni-adjusted p-value from a chi-square test across educational levels, and $p_{adj}$ (trend) the BH-adjusted p-value from a linear regression. [i]Applications and diversity ignore participants who have not used CAIs before to avoid zero-inflation issues (the experience variable captures the relationship).

|  | Overall (N=1636) | Low (n=192) | Middle (n=509) | High (n=427) | Very High (n=508) | $p_{adj}$ (X²) | $p_{adj}$ (trend) |
|---|---|---|---|---|---|---|---|
| *Experience: Yes* | 707 (43.2%) | 54 (28.1%) | 156 (30.6%) | 188 (44.0%) | 309 (60.8%) | <0.001 | 0.000 |
| *Applications (n>0)* | 2.2 (1.4) | 1.8 (1.2) | 1.9 (1.4) | 2.0 (1.1) | 2.7 (1.5) | **<0.001** | **<0.001** |
| *CAI diversity (n>0)* | 1.4 (0.7) | 1.2 (0.5) | 1.2 (0.4) | 1.3 (0.6) | 1.5 (0.9) | **<0.001** | **<0.001** |
| *Behavioural Intention (BI)* | 3.7 (2.0) | 3.0 (1.8) | 3.2 (1.8) | 3.8 (1.9) | 4.4 (2.0) | **<0.001** | **<0.001** |
| *Performance Expectancy* | 4.2 (1.7) | 3.6 (1.7) | 3.8 (1.7) | 4.2 (1.6) | 4.7 (1.7) | **<0.001** | **<0.001** |
| *Effort Expectancy* | 4.3 (1.7) | 3.5 (1.8) | 3.9 (1.7) | 4.5 (1.6) | 4.9 (1.6) | **<0.001** | **<0.001** |
| *Social Influence* | 2.9 (1.9) | 2.5 (1.8) | 2.5 (1.7) | 2.9 (1.9) | 3.6 (2.0) | **<0.001** | **<0.001** |
| *Facilitation Conditions* | 4.3 (1.9) | 3.3 (1.9) | 3.8 (1.9) | 4.5 (1.8) | 5.0 (1.6) | **<0.001** | **<0.001** |
| *Hedonistic Motivation* | 3.8 (1.9) | 3.2 (1.8) | 3.4 (1.8) | 3.9 (1.8) | 4.4 (1.8) | **<0.001** | **<0.001** |
| *Habit* | 2.6 (1.9) | 2.1 (1.6) | 2.2 (1.6) | 2.5 (1.8) | 3.2 (2.1) | **<0.001** | **<0.001** |
| *Trust* | 3.4 (1.8) | 3.1 (1.7) | 3.1 (1.7) | 3.5 (1.7) | 3.7 (1.8) | **<0.001** | **<0.001** |
| *Anxiety* | 3.7 (1.5) | 3.8 (1.6) | 3.7 (1.4) | 3.6 (1.5) | 3.6 (1.5) | 1.000 | 1.000 |
| *Negativity* | 2.9 (1.1) | 3.0 (1.2) | 3.0 (1.1) | 2.9 (1.1) | 2.8 (1.1) | 0.842 | 0.123 |
| *Optimism* | 2.7 (1.1) | 2.5 (1.2) | 2.6 (1.1) | 2.7 (1.1) | 3.1 (1.1) | **<0.001** | **<0.001** |
| *Pessimism* | 3.0 (1.2) | 3.1 (1.2) | 3.1 (1.2) | 2.9 (1.1) | 2.8 (1.2) | **0.008** | **0.001** |
| *Perceived advantages* | 2.5 (1.1) | 2.2 (1.1) | 2.4 (1.1) | 2.6 (1.1) | 2.8 (1.1) | **<0.001** | **<0.001** |
| *Perceived disadvantages* | 2.4 (1.1) | 2.4 (1.1) | 2.4 (1.1) | 2.3 (1.1) | 2.4 (1.1) | 1.000 | 1.000 |

All variables that were associated with education are also associated with CAI avoidance (Table 2), i.e., **H4** is supported for all variables (all *p's* < .020) except for anxiety and disadvantage. Additionally, experience, behavioral intention, performance expectancy, social influence, facilitating conditions, and habit are associated with online search avoidance (all *p's* < .023).

**Table 2. Variable distribution and intervention groups and avoidance behavior.** Each of the three sections in the table corresponds to variables associated with experience, UTAUT2, and affective items. $p_{adj}$ (X²) denotes the Bonferroni-adjusted p-value from a chi-square test comparing avoidance (Yes vs. No). [i]Applications for CAI and CAI diversity ignore participants who have not used CAIs before to avoid zero-inflation issues (the experience variable captures the relationship).

|  | CAI | | | Search | | | Control | | |
|---:|:---:|:---:|:---:|:---:|:---:|:---:|:---:|:---:|:---:|
|  | No (n=374) | Yes (n=360) | $p_{adj}$ ($X^2$) | No (n=307) | Yes (n=137) | $p_{adj}$ ($X^2$) | No (n=381) | Yes (n=77) | $p_{adj}$ ($X^2$) |
| Experience: Yes | 207 (57.5%) | 113 (30.2%) | <0.001 | 157 (51.1%) | 38 (27.7%) | <0.001 | 159 (41.7%) | 33 (42.9%) | 1.000 |
| Applications (n>0) | 2.6 (1.6) | 1.9 (1.2) | **<0.001** | 2.1 (1.3) | 1.9 (1.1) | 1.000 | 2.2 (1.3) | 2.4 (1.8) | 1.000 |
| CAI diversity (n>0) | 1.5 (0.9) | 1.2 (0.5) | **0.001** | 1.3 (0.6) | 1.5 (0.8) | 1.000 | 1.3 (0.5) | 1.6 (1.1) | 1.000 |
| Behavioural Intention (BI) | 4.3 (1.9) | 3.2 (1.9) | **<0.001** | 4.0 (2.0) | 3.2 (2.0) | **0.002** | 3.7 (2.0) | 3.4 (2.1) | 1.000 |
| Performance Expectancy | 4.6 (1.6) | 3.8 (1.7) | **<0.001** | 4.4 (1.6) | 3.7 (1.8) | **0.001** | 4.2 (1.7) | 3.9 (1.8) | 1.000 |
| Effort Expectancy | 4.8 (1.5) | 3.8 (1.7) | **<0.001** | 4.7 (1.6) | 3.7 (1.8) | **<0.001** | 4.4 (1.8) | 4.0 (1.9) | 1.000 |
| Social Influence | 3.2 (2.0) | 2.6 (1.8) | **<0.001** | 3.3 (1.9) | 2.6 (1.9) | **0.023** | 2.8 (1.9) | 3.0 (1.9) | 1.000 |
| Facilitation Conditions | 4.9 (1.6) | 3.7 (1.9) | **<0.001** | 4.8 (1.7) | 3.5 (2.0) | **<0.001** | 4.3 (1.9) | 3.8 (1.9) | 0.569 |
| Hedonistic Motivation | 4.3 (1.8) | 3.4 (1.8) | **<0.001** | 4.1 (1.9) | 3.4 (1.9) | **0.013** | 3.8 (1.9) | 3.6 (1.9) | 1.000 |
| Habit | 3.0 (2.0) | 2.1 (1.6) | **<0.001** | 2.7 (1.9) | 2.3 (1.9) | 0.790 | 2.5 (1.8) | 2.8 (2.1) | 1.000 |
| Trust | 3.8 (1.8) | 3.1 (1.7) | **<0.001** | 3.5 (1.7) | 3.1 (1.8) | 0.349 | 3.4 (1.7) | 3.2 (1.9) | 1.000 |
| Anxiety | 3.6 (1.5) | 3.7 (1.4) | 1.000 | 3.5 (1.5) | 3.7 (1.6) | 1.000 | 3.7 (1.5) | 3.8 (1.7) | 1.000 |
| Negativity | 2.8 (1.1) | 3.0 (1.1) | **0.020** | 2.8 (1.1) | 3.0 (1.2) | 0.768 | 2.9 (1.1) | 2.8 (1.2) | 1.000 |
| Optimism | 3.0 (1.1) | 2.5 (1.1) | **<0.001** | 2.9 (1.1) | 2.7 (1.2) | 1.000 | 2.7 (1.1) | 2.8 (1.3) | 1.000 |
| Pessimism | 2.8 (1.2) | 3.2 (1.1) | **0.003** | 2.8 (1.1) | 3.0 (1.3) | 1.000 | 2.9 (1.2) | 2.9 (1.2) | 1.000 |
| Perceived advantages | 2.7 (1.1) | 2.3 (1.0) | **<0.001** | 2.6 (1.1) | 2.6 (1.1) | 1.000 | 2.4 (1.1) | 2.7 (1.3) | 1.000 |
| Perceived disadvantages | 2.4 (1.1) | 2.4 (1.1) | 1.000 | 2.3 (1.1) | 2.6 (1.2) | 0.247 | 2.3 (1.1) | 2.4 (1.2) | 1.000 |

Figure 2 illustrates how the variables positively associated with education are also negatively associated with CAI avoidance and vice versa. For example, the higher educational attainments, the more experience with CAIs, and the more experience with CAIs, the less likely participants were to avoid the task. It also illustrates a similar phenomenon for age and, to a much lesser degree, gender. This pattern is also evident in online search avoidance behavior, although the associations are weaker than the ones for CAI avoidance.

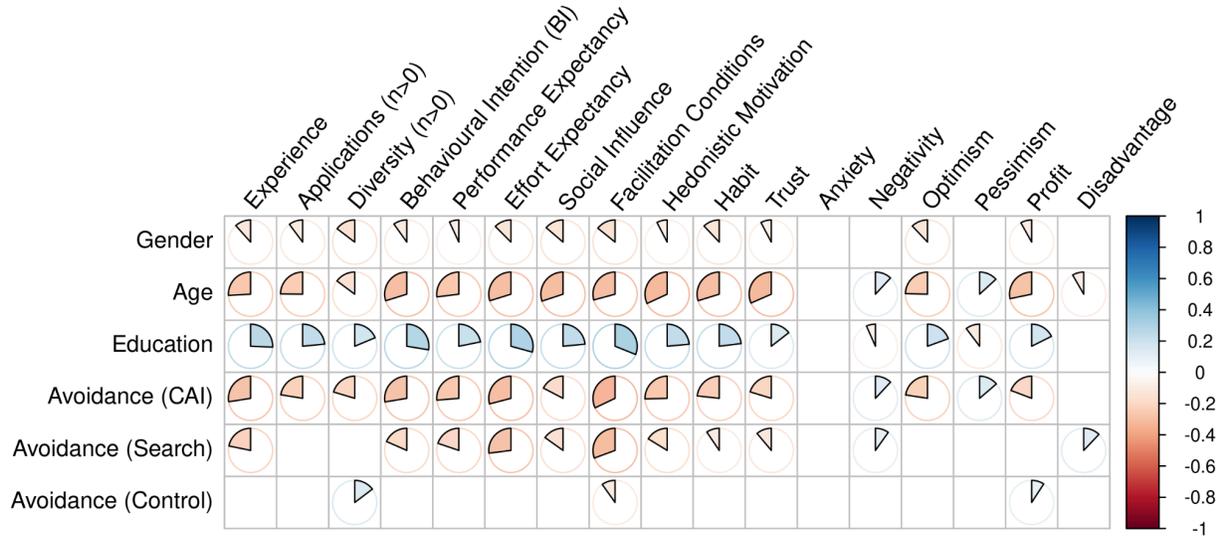

**Figure 2. Pearson correlation between variables and socio-demographics.** Education is linearized (low < middle < high < very high). Avoidance correlations (last three rows) are calculated within each treatment group. Only statistically significant correlations are shown (p < .05).

To understand the role of education in the presence of the UTAUT2 variables, we employ a structural equation model approach (Figure 3) where we show that education is significantly associated with avoidance in the CAI group (**H5**). The magnitude of the education point estimate (.23) is comparable to established moderators of the UTAUT: age (.32), gender (.22), and experience (.20). All diagnostic metrics indicated an acceptable fit according to cut-off values (Kline, 2016): CFI>.9, TLI>.9, RMSEA < .08, SRMR < .08, and Rel. X2 < 3. The variance explained (R2) for the behavioral intention and avoidance are .79 and .35, respectively. Cronbach's alphas are > 0.9 for multi-item UTAUT2 constructs (Appendix E).

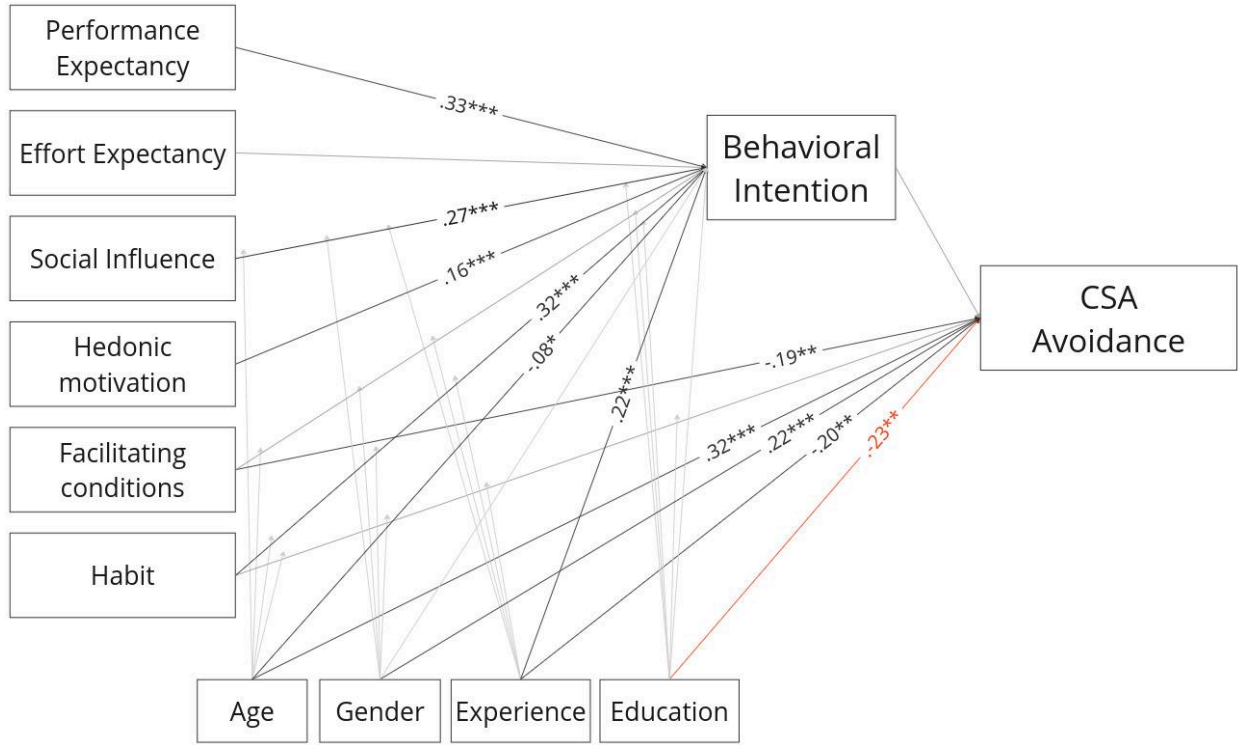

**Figure 3. Structural equation model estimates for a UTAUT2 model and Education on CAI avoidance.** We incorporated educational levels as a predictor of behavioral intention and CAI avoidance in the original UTAUT2 model. We also included interaction terms (with education) following existent interactions with demographics in the original model. Darker lines represent statistically significant relationships.

As an additional robustness check, we report that the education point estimate is consistent across variations of UTAUT2 models, regardless of weights (Appendix D1), absence or presence of interaction effects (Appendix D1 and D2), absence or presence of the experience variable (as, per H2, more educated individuals might also have more likely previous experience with CAIs) (Appendix D2), absence or presence of facilitating conditions and habit (Appendix D2 and D3), removing the behavioral intention (BI) path (Appendix D4) and including factor loadings for single-item variables (Appendix D4). Focusing only on the BI path (Appendix D5), i.e., BI as a dependent variable (without CAI avoidance), we find statistically significant effects consistent with Figure 3 (but with different magnitudes for the estimates), except for education, which is now significant predictor of BI (.05, $p<.001$) and age, which is not a significant predictor of BI. In line with Venkatesh et al. (2012), BI is only a significant predictor of avoidance itself if we remove UTAUT2 latent variables as predictors of avoidance (Appendix D3).

LASSO regression with 10-fold cross-validation was applied to identify associations between demographic variables, affective CAI variables, and experience, controlling for attentiveness and the timer. Our results support **H6**: education was consistently selected by the LASSO regression and maintained statistical significance in subsequent logistic regression models (Table 3). Other demographic variables, namely gender, age, and experience, also emerged as robust predictors of CAI avoidance. Although the LASSO procedure also identified several affective variables, these did not achieve statistical significance in the logistic regression analyses.

**Table 3. LASSO and logistic regression on CAI avoidance.** Results from LASSO regression with 10-fold cross-validation and corresponding logistic regression models. LASSO was used to select variables, while logistic regression provided more interpretable coefficient estimates to assess robustness. The table includes variable selection using the optimal regularization parameter that minimizes the mean squared error ($\lambda_{min}$) and the more parsimonious model corresponding to the largest lambda value within one standard error of the minimum error ($\lambda_{1se}$). See Appendix F for other weights.

|  | Raked and Trimmed | | | |
| --- | --- | --- | --- | --- |
|  | $\lambda_{min}$ | | $\lambda_{1se}$ | |
|  | LASSO | GLM | LASSO | GLM |
| Gender | 0.06 | 0.173** | 0.03 | 0.190** |
| Age | 0.09 | 0.241*** | 0.07 | 0.240*** |
| Education | -0.07 | -0.181** | -0.05 | -0.182** |
| Experience | -0.07 | -0.226** | -0.06 | -0.223** |
| Trust |  |  |  |  |
| Anxiety |  |  |  |  |
| Negativity | 0.01 | 0.034 | 0.00 | 0.030 |
| Optimism | -0.02 | -0.042 | -0.01 | -0.060 |
| Pessimism | 0.02 | 0.054 | 0.01 | 0.036 |
| Perceived advantages |  |  |  |  |
| Perceived disadvantages | 0.01 | -0.033 |  |  |
| Attentiveness | -0.04 | -0.061 | -0.00 | -0.069 |
| Timer | -0.02 | -0.163* |  |  |

## Discussion

Results across multiple analytical approaches consistently support hypotheses regarding task avoidance in response to different intervention types and education. Avoidance rates differed markedly across groups, with CAI eliciting significantly higher avoidance than the search and control conditions (H1). We found significant interactions between demographic variables and CAI avoidance, with lower education associated with higher avoidance (H2). In line with H3 and H4, we observed that higher education was positively associated with UTAUT2 constructs, CAI experience, and favorable affective attitudes and that these variables were, in turn, associated with lower avoidance. Structural equation modeling further confirmed the central role of education in shaping behavioral intentions and avoidance within the CAI context (H5), with education explaining variance comparable to established UTAUT2 socio-demographic predictors. Similarly, LASSO regression consistently identified education as a predictor of CAI avoidance (H6), reinforcing its role in shaping engagement with AI, even when avoidance entailed forfeiting a monetary reward. Our study focused on individuals aged 25 to 64, a cohort with largely completed formal education and generally active in the labor market, often participating in online studies to supplement their income (Keusch, 2015).

While task avoidance was present in both the CAI and search groups, it was notably higher in the CAI group despite that task being simpler: the search task involved more procedural demands, as participants had to navigate to a search engine, manage browser tabs, choose relevant links, and copy a resource, requiring more effort and self-direction. Given that our sample consisted of online panelists likely familiar with similar tasks, we interpret the higher avoidance in the CAI condition not as a reflection of task difficulty but as an expression of their reluctance to use CAI specifically. This is particularly striking given that the CAI task required minimal action (clicking a link, entering a prompt, and copying a link) and was accompanied by visually-aided instructions. The mechanisms behind this reluctance to engage with the task could be explained by the UTAUT2 latent and affective variables as suggested by generally strong correlations. For example, perceived task difficulty is captured by effort expectancy, which, as per the original UTAUT2 specification and our modeling, is not theorized to influence task avoidance directly but behavioral intention (Venkatesh et al., 2012). Alternatively, in a recent systematic review, Leschanowsky et al. (2024) suggested that reluctance around CAI may also stem from entangled perceptions of privacy, security, and trust—conceptually overlapping constructs yet rarely studied together in the CAI context. Avoidance may reflect a resistance rooted in broader trust-related dynamics, extending beyond the technology itself to include concerns about the companies and politics behind it.

The linear downward relationship of the role of education with avoidance in the search and CAI groups implies a general technology avoidance from lower educated populations, though the increased elevation for CAI participants suggests an additional mechanism, likely tied to the novelty of the technology, where education may play a larger role in shaping participants' willingness or confidence to engage. While our results show a strong relationship between education and CAI avoidance, the relationship between education and behavioral intention (BI) is only significant in UTAUT2 models that exclude the behavioral path. In other words, once individuals are confronted with a behavioral task, the importance of education on intention appears to be negligible compared to its direct role in the behavior.

We have centered our analysis on education to contribute to a more nuanced and actionable understanding of user diversity in digital adoption. A parallel narrative could have been constructed focusing on age, arriving at similar conclusions to those we reached for education. The role of gender, by contrast, proved more nuanced as we only found it interacted with the CAI condition but not the search one. Both of them remained significant in both our SEM-UTAUT2 and LASSO analyses. Crucially, the observed patterns for age and gender align with previous findings in digital inequality research, lending additional support to the validity of our analyses.

Overall, our study provides evidence for the role of education by employing a behavioral metric within an experimental design, moving beyond the predominant reliance on self-reported or correlational data in the existing literature. Accordingly, our findings support the inclusion of education as a key variable within the UTAUT2 framework within the genAI context.

The uneven engagement with CAI that we observe raises concerns about a broader societal impact, namely, the risk of exacerbating existing inequalities if certain groups systematically opt-out and are underrepresented in CAI use and research. Studies involving CAI or similar technologies may be susceptible to self-selection bias in which those in an already advantageous position are more likely to participate, leading to conclusions that overlook the perspectives and needs of disadvantaged groups. Recognizing and addressing this dynamic is crucial for developing equitable and inclusive digital systems.

**Limitations**

The analysis and results presented here are based on an analysis conducted due to high attrition rates in the CAI treatment observed in the context of another research project. The panel agency informed us about the issue early in the data collation, prompting us to remove the time restriction for this group to enhance participation. This adjustment resulted in a heterogeneous CAI treatment, where some participants completed the task with a timer while others did not. While our analyses reveal that removing the timer did not reduce avoidance, this should be considered when interpreting the results.

Secondly, though all treatment groups were concurrently filled as participants entered the study over time, they were not perfectly balanced regarding demographics. This imbalance may have arisen either by chance or due to a complex interaction between socio-demographic differences in the timing of survey participation, higher duration in the CAI and search tasks, and higher CAI and search abandonment of the survey during the task. To address the imbalances, we applied weighting procedures to align our sample distributions with the German general population, which has been shown to help mitigate biases when attempting to represent an entire population with an online sample (Grewenig et al., 2023). This, together with the consistency of results across weighing strategies, provides a fair degree of confidence. Still, we note that weights were calculated using only three socio-demographic variables and caution against generalizations beyond the corresponding context of the study, i.e., avoidance of a single CAI-related task with a German online sample. Regarding generalizations off the online panelists, they tend to be more tech-savvy and financially motivated than the general population, and their behavior may not reflect that of less digitally engaged or less incentivized individuals.

Finally, although the CAI task was designed to be simpler than the search task, we did not measure perceived or experienced task difficulty, usability, or frustration directly in relation to the avoidance, which could help disentangle avoidance due to UTAUT factors, low motivation, or other user experience issues; that said, technical feasibility was pre-tested and regularly checked throughout the study duration, and users had the opportunity to contact the researchers or Bilendi about technical issues, and we did not receive any reports in this direction.

**Conclusion**

This study examined how educational attainment relates to avoidance of conversational AI in an experimental setting. Across multiple analytical approaches, we observed that lower education was associated with a higher likelihood of avoiding a task involving CAI, supported by associations across UTAUT and affective factors. The consistency of this association across methods suggests that educational differences may play a role in shaping initial engagement with emerging AI tools. These findings highlight the value of including education in theoretical frameworks of technology adoption and raise considerations for designing inclusive digital systems.

.

**References**

Abu-Shanab, E. A. (2011). Education level as a technology adoption moderator. *2011 3rd International Conference on Computer Research and Development*, 324–328. https://doi.org/10.1109/iccrd.2011.5764029

Alsharhan, A., Al-Emran, M., & Shaalan, K. (2024). Chatbot Adoption: A Multiperspective Systematic Review and Future Research Agenda. *IEEE Transactions on Engineering Management*, *71*, 10232–10244. IEEE Transactions on Engineering Management. https://doi.org/10.1109/TEM.2023.3298360

Araujo, T., Helberger, N., Kruikemeier, S., & de Vreese, C. H. (2020). In AI we trust? Perceptions about automated decision-making by artificial intelligence. *AI & SOCIETY*, *35*(3), 611–623. https://doi.org/10.1007/s00146-019-00931-w

Austin, P. C. (2009). Balance diagnostics for comparing the distribution of baseline covariates between treatment groups in propensity-score matched samples. *Statistics in Medicine*, *28*(25), 3083–3107. https://doi.org/10.1002/sim.3697

Bao, L., Krause, N. M., Calice, M. N., Scheufele, D. A., Wirz, C. D., Brossard, D., Newman, T. P., & Xenos, M. A. (2022). Whose AI? How different publics think about AI and its social impacts. *Computers in Human Behavior*, *130*, 107182. https://doi.org/10.1016/j.chb.2022.107182

Beauducel, A., & Herzberg, P. Y. (2006). On the Performance of Maximum Likelihood Versus Means and Variance Adjusted Weighted Least Squares Estimation in CFA. *Structural Equation Modeling: A Multidisciplinary Journal*, *13*(2), 186–203. https://doi.org/10.1207/s15328007sem1302_2

Bentley, S. V., Naughtin, C. K., McGrath, M. J., Irons, J. L., & Cooper, P. S. (2024). The digital

divide in action: How experiences of digital technology shape future relationships with artificial intelligence. *AI and Ethics*, *4*(4), 901–915. https://doi.org/10.1007/s43681-024-00452-3

Biswas, M., & Murray, J. (2024). The Impact of Education Level on AI Reliance, Habit Formation, and Usage. *2024 29th International Conference on Automation and Computing (ICAC)*, 1–6. https://doi.org/10.1109/ICAC61394.2024.10718860

Budhathoki, T., Zirar, A., Njoya, E. T., & Timsina, A. (2024). ChatGPT adoption and anxiety: A cross-country analysis utilising the unified theory of acceptance and use of technology (UTAUT). *Studies in Higher Education*, *49*(5), 831–846. https://doi.org/10.1080/03075079.2024.2333937

Capraro, V., Lentsch, A., Acemoglu, D., Akgun, S., Akhmedova, A., Bilancini, E., Bonnefon, J.-F., Brañas-Garza, P., Butera, L., Douglas, K. M., Everett, J. A. C., Gigerenzer, G., Greenhow, C., Hashimoto, D. A., Holt-Lunstad, J., Jetten, J., Johnson, S., Kunz, W. H., Longoni, C., … Viale, R. (2024). The impact of generative artificial intelligence on socioeconomic inequalities and policy making. *PNAS Nexus*, *3*(6), pgae191. https://doi.org/10.1093/pnasnexus/pgae191

Casheekar, A., Lahiri, A., Rath, K., Prabhakar, K. S., & Srinivasan, K. (2024). A contemporary review on chatbots, AI-powered virtual conversational agents, ChatGPT: Applications, open challenges and future research directions. *Computer Science Review*, *52*, 100632. https://doi.org/10.1016/j.cosrev.2024.100632

Cruz-Jesus, F., Vicente, M. R., Bacao, F., & Oliveira, T. (2016). The education-related digital divide: An analysis for the EU-28. *Computers in Human Behavior*, *56*, 72–82. https://doi.org/10.1016/j.chb.2015.11.027

Daepp, M. I. G., & Counts, S. (2024). *The Emerging AI Divide in the United States* (No. arXiv:2404.11988). arXiv. https://doi.org/10.48550/arXiv.2404.11988

Dubiel, M., Halvey, M., Azzopardi, L., & Daronnat, S. (2018, July 12). *Investigating how conversational search agents affect user's behaviour, performance and search experience: The Second International Workshop on Conversational Approaches to Information Retrieval*. https://sites.google.com/view/cair-ws/cair-2018

Eder, M., & Sjøvaag, H. (2024). Artificial intelligence and the dawn of an algorithmic divide. *Frontiers in Communication*, *9*. https://doi.org/10.3389/fcomm.2024.1453251

Elena-Bucea, A., Cruz-Jesus, F., Oliveira, T., & Coelho, P. S. (2021). Assessing the Role of Age, Education, Gender and Income on the Digital Divide: Evidence for the European Union. *Information Systems Frontiers*, *23*(4), 1007–1021. https://doi.org/10.1007/s10796-020-10012-9

Goldenthal, E., Park, J., Liu, S. X., Mieczkowski, H., & Hancock, J. T. (2021). Not All AI are Equal: Exploring the Accessibility of AI-Mediated Communication Technology. *Computers in Human Behavior*, *125*, 106975. https://doi.org/10.1016/j.chb.2021.106975

Grewenig, E., Lergetporer, P., Simon, L., Werner, K., & Woessmann, L. (2023). Can internet surveys represent the entire population? A practitioners' analysis. *European Journal of Political Economy*, *78*, 102382. https://doi.org/10.1016/j.ejpoleco.2023.102382

Hew, J.-J., Lee, V.-H., Ooi, K.-B., & Wei, J. (2015). What catalyses mobile apps usage intention: An empirical analysis. *Industrial Management & Data Systems*, *115*(7), 1269–1291. https://doi.org/10.1108/IMDS-01-2015-0028

Hong, J.-W. (2022). I Was Born to Love AI: The Influence of Social Status on AI Self-Efficacy and Intentions to Use AI. *International Journal of Communication*, *16*(0), Article 0.

Kabalisa, R., & Altmann, J. (2021). AI Technologies and Motives for AI Adoption by Countries and Firms: A Systematic Literature Review. In K. Tserpes, J. Altmann, J. Á. Bañares, O. Agmon Ben-Yehuda, K. Djemame, V. Stankovski, & B. Tuffin (Eds.), *Economics of Grids, Clouds, Systems, and Services* (pp. 39–51). Springer International Publishing. https://doi.org/10.1007/978-3-030-92916-9_4

Kacperski, C., Ulloa, R., Bonnay, D., Kulshrestha, J., Selb, P., & Spitz, A. (2025). Characteristics of ChatGPT users from Germany: Implications for the digital divide from web tracking data. *PLOS ONE*, *20*(1), e0309047. https://doi.org/10.1371/journal.pone.0309047

Kaushik, A., & Jones, G. J. F. (2023). Comparing Conventional and Conversational Search Interaction using Implicit Evaluation Methods. *Proceedings of the 18th International Joint Conference on Computer Vision, Imaging and Computer Graphics Theory and Applications*, 292–304. https://doi.org/10.5220/0011798500003417

Kelly, S., Kaye, S.-A., & Oviedo-Trespalacios, O. (2023). What factors contribute to the acceptance of artificial intelligence? A systematic review. *Telematics and Informatics*, *77*, 101925. https://doi.org/10.1016/j.tele.2022.101925

Keusch, F. (2015). Why do people participate in Web surveys? Applying survey participation theory to Internet survey data collection. *Management Review Quarterly*, *65*(3), 183–216. https://doi.org/10.1007/s11301-014-0111-y

Khowaja, S. A., Khuwaja, P., Dev, K., Wang, W., & Nkenyereye, L. (2024). ChatGPT Needs SPADE (Sustainability, PrivAcy, Digital divide, and Ethics) Evaluation: A Review. *Cognitive Computation*. https://doi.org/10.1007/s12559-024-10285-1

Kline, R. B. (2016). *Principles and practice of structural equation modeling* (4th ed). Guilford.

Lenth, R. V. (2024). *emmeans: Estimated Marginal Means, aka Least-Squares Means*.

https://CRAN.R-project.org/package=emmeans

Leschanowsky, A., Rech, S., Popp, B., & Bäckström, T. (2024). Evaluating privacy, security, and trust perceptions in conversational AI: A systematic review. *Computers in Human Behavior*, *159*, 108344. https://doi.org/10.1016/j.chb.2024.108344

Li, Q., Yang, D., & Chen, X. (2014). Predicting Determinants and Moderating Factors of Mobile Phone Data Flow Service Adoption. *2014 Seventh International Joint Conference on Computational Sciences and Optimization*, 390–394. https://doi.org/10.1109/CSO.2014.82

Liew, E. J. Y., Vaithilingam, S., & Nair, M. (2014). Facebook and socio-economic benefits in the developing world. *Behaviour & Information Technology*, *33*(4), 345–360. https://doi.org/10.1080/0144929x.2013.810775

Lythreatis, S., Singh, S. K., & El-Kassar, A.-N. (2022). The digital divide: A review and future research agenda. *Technological Forecasting and Social Change*, *175*, 121359. https://doi.org/10.1016/j.techfore.2021.121359

Ma, X., & Huo, Y. (2023). Are users willing to embrace ChatGPT? Exploring the factors on the acceptance of chatbots from the perspective of AIDUA framework. *Technology in Society*, *75*, 102362. https://doi.org/10.1016/j.techsoc.2023.102362

McClain, C. (2024, March 26). Americans' use of ChatGPT is ticking up, but few trust its election information. *Pew Research Center*. https://www.pewresearch.org/short-reads/2024/03/26/americans-use-of-chatgpt-is-ticking-up-but-few-trust-its-election-information/

Nguyen, M. H., Hargittai, E., & Marler, W. (2021). Digital inequality in communication during a time of physical distancing: The case of COVID-19. *Computers in Human Behavior*, *120*,

106717. https://doi.org/10.1016/j.chb.2021.106717

Niehaves, B., & Plattfaut, R. (2010). The Age-Divide in Private Internet Usage: A Quantitative Study of Technology Acceptance. *AMCIS 2010 Proceedings*. https://aisel.aisnet.org/amcis2010/407

Perrig, S., Scharowski, N., & Brühlmann, F. (2023). Trust Issues with Trust Scales: Examining the Psychometric Quality of Trust Measures in the Context of AI. *Extended Abstracts of the 2023 CHI Conference on Human Factors in Computing Systems*, 1–7. https://doi.org/10.1145/3544549.3585808

Radhakrishnan, J., & Chattopadhyay, M. (2020). Determinants and Barriers of Artificial Intelligence Adoption – A Literature Review. In S. K. Sharma, Y. K. Dwivedi, B. Metri, & N. P. Rana (Eds.), *Re-imagining Diffusion and Adoption of Information Technology and Systems: A Continuing Conversation* (pp. 89–99). Springer International Publishing. https://doi.org/10.1007/978-3-030-64849-7_9

Radlinski, F., & Craswell, N. (2017). A Theoretical Framework for Conversational Search. *Proceedings of the 2017 Conference on Conference Human Information Interaction and Retrieval*, 117–126. https://doi.org/10.1145/3020165.3020183

Rosseel, Y. (2012). lavaan: An R Package for Structural Equation Modeling. *Journal of Statistical Software*, *48*(2), 1–36. https://doi.org/10.18637/jss.v048.i02

Said, N., Potinteu, A. E., Brich, I., Buder, J., Schumm, H., & Huff, M. (2023). An artificial intelligence perspective: How knowledge and confidence shape risk and benefit perception. *Computers in Human Behavior*, *149*, 107855. https://doi.org/10.1016/j.chb.2023.107855

Silber, H., Stadtmüller, S., & Cernat, A. (2023). Comparing participation motives of professional

and non-professional respondents. *International Journal of Market Research*, *65*(4), 361–372. https://doi.org/10.1177/14707853231166882

Statistisches Bundesamt. (2024). *Destatis*. Statistisches Bundesamt. https://www.destatis.de/DE/Home/_inhalt.html

Stöhr, C., Ou, A. W., & Malmström, H. (2024). Perceptions and usage of AI chatbots among students in higher education across genders, academic levels and fields of study. *Computers and Education: Artificial Intelligence*, *7*, 100259. https://doi.org/10.1016/j.caeai.2024.100259

Stokel-Walker, C., & Van Noorden, R. (2023). What ChatGPT and generative AI mean for science. *Nature*, *614*(7947), 214–216. https://doi.org/10.1038/d41586-023-00340-6

Tavitiyaman, P., Zhang, X., & Tsang, W. Y. (2022). How Tourists Perceive the Usefulness of Technology Adoption in Hotels: Interaction Effect of Past Experience and Education Level. *Journal of China Tourism Research*, *18*(1), 64–87. https://doi.org/10.1080/19388160.2020.1801546

van Dijk, J. A. G. M., Peters, O., & Ebbers, W. (2008). Explaining the acceptance and use of government Internet services: A multivariate analysis of 2006 survey data in the Netherlands. *Government Information Quarterly*, *25*(3), 379–399. https://doi.org/10.1016/j.giq.2007.09.006

Venkatesh, V. (2022). Adoption and use of AI tools: A research agenda grounded in UTAUT. *Annals of Operations Research*, *308*(1), 641–652. https://doi.org/10.1007/s10479-020-03918-9

Venkatesh, V., Morris, M. G., Davis, G. B., & Davis, F. D. (2003). User Acceptance of Information Technology: Toward a Unified View. *MIS Quarterly*, *27*(3), 425–478.

https://doi.org/10.2307/30036540

Venkatesh, V., Thong, J. Y. L., & Xu, X. (2012). Consumer Acceptance and Use of Information Technology: Extending the Unified Theory of Acceptance and Use of Technology. *MIS Quarterly*, *36*(1), 157–178. https://doi.org/10.2307/41410412

Venkatesh, V., Thong, J. Y. L., & Xu, X. (2016). *Unified Theory of Acceptance and Use of Technology: A Synthesis and the Road Ahead* (SSRN Scholarly Paper No. 2800121). https://papers.ssrn.com/abstract=2800121

Vimalkumar, M., Sharma, S. K., Singh, J. B., & Dwivedi, Y. K. (2021). 'Okay google, what about my privacy?': User's privacy perceptions and acceptance of voice based digital assistants. *Computers in Human Behavior*, *120*, 106763. https://doi.org/10.1016/j.chb.2021.106763

Wang, C., Boerman, S. C., Kroon, A. C., Möller, J., & H de Vreese, C. (2024). The artificial intelligence divide: Who is the most vulnerable? *New Media & Society*, 14614448241232345. https://doi.org/10.1177/14614448241232345

Wang, Y.-Y., & Wang, Y.-S. (2022). Development and validation of an artificial intelligence anxiety scale: An initial application in predicting motivated learning behavior. *Interactive Learning Environments*, *30*(4), 619–634. https://doi.org/10.1080/10494820.2019.1674887

Zhang, B., & Dafoe, A. (2019). Artificial Intelligence: American Attitudes and Trends. *SSRN Electronic Journal*. https://doi.org/10.2139/ssrn.3312874

# Appendix

## A. Distribution of participants across groups and demographic variables

**Distribution of participants across groups and demographic variables.** The first and second columns indicate, respectively, the name of the demographic and the distribution in the German population (2023). The next four columns show the distribution across demographics and groups. An absolute standardized mean difference (SMD) below 0.1 generally indicates an acceptable covariate balance between treatment groups (Austin, 2009). The p-values show significant differences between each of the groups and across the three groups.

|  | German Pop. | [ALL] | Control | Search | CAI | SD | p-values Control vs. Search | p-values Control vs. CAI | p-values Search vs. CAI | Overall |
|---|---|---|---|---|---|---|---|---|---|---|
| Gender |  |  |  |  |  | 0.094 | 0.006 | 0.019 | 0.480 | 0.011 |
|   Women | 49.7% | 780 (47.7%) | 192 (41.9%) | 228 (51.4%) | 360 (49.0%) |  |  |  |  |  |
|   Men | 50.3% | 856 (52.3%) | 266 (58.1%) | 216 (48.6%) | 374 (51.0%) |  |  |  |  |  |
| Age |  | 46.8 (12.0) | 46.7 (11.4) | 46.7 (11.9) | 46.9 (12.3) | 0.015 | 0.968 | 0.812 | 0.853 | 0.968 |
| AgeCat |  |  |  |  |  | 0.035 | 0.108 | 0.007 | 0.661 | 0.034 |
|   25-35 | 24.7% | 329 (20.1%) | 84 (18.3%) | 85 (19.1%) | 160 (21.8%) |  |  |  |  |  |
|   35-45 | 25.1% | 392 (24.0%) | 112 (24.5%) | 111 (25.0%) | 169 (23.0%) | 0.020 |  |  |  |  |
|   45-55 | 23% | 357 (21.8%) | 124 (27.1%) | 91 (20.5%) | 142 (19.3%) | 0.077 |  |  |  |  |
|   55-65 | 27.2% | 558 (34.1%) | 138 (30.1%) | 157 (35.4%) | 263 (35.8%) | 0.057 |  |  |  |  |
| Education |  |  |  |  |  | 0.023 | 0.013 | 0.066 | 0.773 | 0.068 |
|   Low | 18.9% | 192 (11.7%) | 61 (13.3%) | 49 (11.0%) | 82 (11.2%) |  |  |  |  |  |
|   Middle | 25.7% | 509 (31.1%) | 152 (33.2%) | 129 (29.1%) | 228 (31.1%) | 0.041 |  |  |  |  |
|   High | 35.1% | 427 (26.1%) | 94 (20.5%) | 133 (30.0%) | 200 (27.2%) | 0.094 |  |  |  |  |
|   Very High | 20.3% | 508 (31.1%) | 151 (33.0%) | 133 (30.0%) | 224 (30.5%) | 0.030 |  |  |  |  |
| Attentiveness |  | 0.9 (0.2) | 1.0 (0.2) | 0.9 (0.2) | 0.9 (0.2) | 0.035 | 0.603 | 0.805 | 0.738 | 0.871 |

## B. Weight calculation and diagnostics

**Table #D#1 Diagnostic statistics of the resulting weights.** The first column indicates the type of weight, followed by the intervention group (including overall), the number of weights above 5 standard deviations, the number of times of the maximum weight over the Mdn, the Coefficient of Variance (CV), the Effective Sample Size (ESS), the actual size of the sample (N) and the Kish's Design Effect Due to Weighting (DEffK).

| Weight | Group | N(Weights > 5*Mdn) | Max (times of weight over Mdn) | CV | ESS | N | DEffK |
|---|---|---|---|---|---|---|---|
| Post-Stratification | Overall | 3 | 11.61 | 0.59 | 1,218.08 | 1,636 | 1.34 |
|  | Control | 1 | 12.31 | 0.73 | 299.77 | 458 | 1.53 |
|  | Search | 2 | 6.05 | 0.57 | 334.26 | 444 | 1.33 |
|  | CAI | 0 | 3.53 | 0.48 | 594.94 | 734 | 1.23 |
| Raked | Overall | 0 | 3.88 | 0.53 | 1,282.20 | 1,636 | 1.28 |
|  | Control | 0 | 3.52 | 0.53 | 358.71 | 458 | 1.28 |
|  | Search | 0 | 2.93 | 0.40 | 382.29 | 444 | 1.16 |
|  | CAI | 0 | 2.73 | 0.43 | 618.98 | 734 | 1.19 |
| Trimmed | Overall | 0 | 2.93 | 0.50 | 1,306.00 | 1,636 | 1.25 |
|  | Control | 0 | 2.67 | 0.48 | 372.58 | 458 | 1.23 |
|  | Search | 0 | 2.31 | 0.39 | 385.84 | 444 | 1.15 |
|  | CAI | 0 | 2.73 | 0.43 | 618.98 | 734 | 1.19 |

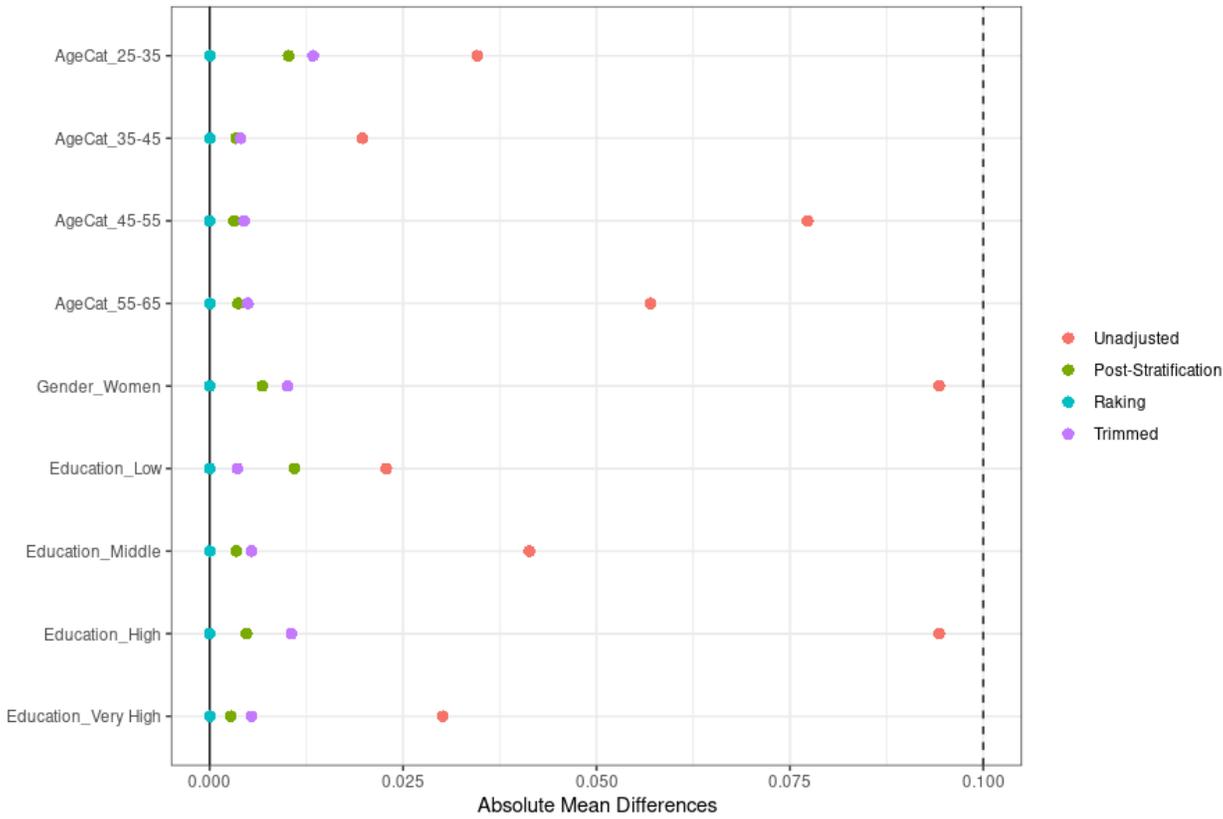

**Figure #D#1. Absolute standardized mean differences.** The plot displays the absolute standardized mean differences (ASMDs) for each gender, age category, and education before (unadjusted) and after applying the post-stratification, raking, and trimming weighting procedure. A vertical dashed line at 0.1 denotes the commonly accepted threshold for acceptable balance.

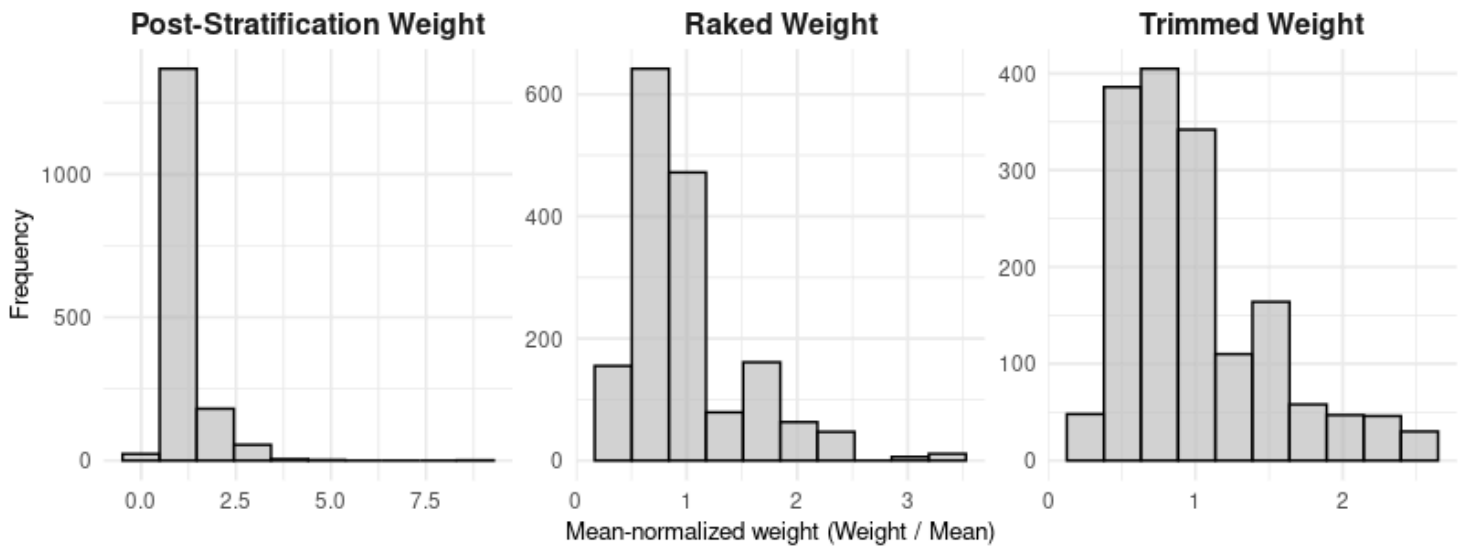

**Figure #D#2. Histograms of weight distributions.** The histograms show in the X-axis the mean-normalized weight. The (right) skewness indicates extreme over-representation of certain individuals, potentially inducing biases towards them.

# C. Full logistic regressions on avoidance

## C1. Effects of intervention on avoidance

**Logistic regression of intervention on avoidance (Search as Reference group)** Odds ratios correspond to a logistic regression fitting the intervention group (IV) on the avoidance (DV). The Control group is used as a reference group.

| Predictors | Unweighted Odds Ratios | p | Raked and Trimmed Odds Ratios | p | Post-stratification Odds Ratios | p |
|---|---|---|---|---|---|---|
| (Intercept) | 0.20 (0.16 – 0.26) | **<0.001** | 0.21 (0.16 – 0.27) | **<0.001** | 0.18 (0.14 – 0.25) | **<0.001** |
| Group [Search] | 2.21 (1.61 – 3.04) | **<0.001** | 2.25 (1.59 – 3.19) | **<0.001** | 2.37 (1.65 – 3.42) | **<0.001** |
| Group [CAI] | 5.14 (3.89 – 6.87) | **<0.001** | 5.16 (3.77 – 7.07) | **<0.001** | 5.76 (4.16 – 8.00) | **<0.001** |
| Observations | 1636 | | 1636 | | 1636 | |
| R2 Tjur | 0.091 | | 0.072 / 0.071 | | 0.082 / 0.081 | |

**Logistic regression of intervention on avoidance (Control as Reference group)** Odds ratios correspond to a logistic regression fitting the intervention group (IV) on the avoidance (DV). The Control group is used as a reference group.

| Predictors | Unweighted Odds Ratios | p | Raked and Trimmed Odds Ratios | p | Post-stratification Odds Ratios | p |
|---|---|---|---|---|---|---|
| (Intercept) | 0.45 (0.36 – 0.54) | <0.001 | 0.46 (0.37 – 0.58) | <0.001 | 0.44 (0.35 – 0.55) | **<0.001** |
| Group [Control] | 0.45 (0.33 – 0.62) | <0.001 | 0.44 (0.31 – 0.63) | <0.001 | 0.42 (0.29 – 0.61) | **<0.001** |
| Group [CAI] | 2.33 (1.82 – 2.99) | <0.001 | 2.29 (1.75 – 2.99) | <0.001 | 2.43 (1.84 – 3.21) | **<0.001** |
| Observations | 1636 | | 1636 | | 1636 | |
| R2 Tjur | 0.091 | | 0.072 / 0.071 | | 0.082 / 0.081 | |

**Logistic regression of intervention on avoidance (Control as Reference group)** Odds ratios correspond to a logistic regression fitting the intervention group (IV) on the avoidance (DV). The Control group is used as a reference group. It includes Time interacting with th groups to test if the timer had an effect reducing avoidance

| Predictors | Unweighted Odds Ratios | p | Raked and Trimmed Odds Ratios | p | Post-stratification Odds Ratios | p |
|---|---|---|---|---|---|---|
| (Intercept) | 0.18 (0.13 – 0.25) | <0.001 | 0.19 (0.13 – 0.28) | <0.001 | 0.17 (0.12 – 0.25) | <0.001 |
| Group [Search] | 2.55 (1.68 – 3.91) | <0.001 | 2.39 (1.51 – 3.78) | <0.001 | 2.51 (1.55 – 4.06) | <0.001 |
| Group [CAI] | 6.59 (4.54 – 9.75) | <0.001 | 6.20 (4.07 – 9.45) | <0.001 | 6.93 (4.50 – 10.67) | <0.001 |
| Timer [with] | 1.28 (0.78 – 2.09) | 0.325 | 1.19 (0.69 – 2.05) | 0.538 | 1.20 (0.68 – 2.14) | 0.527 |
| Group [Search] × Timer [with] | 0.71 (0.38 – 1.36) | 0.306 | 0.87 (0.43 – 1.76) | 0.698 | 0.88 (0.42 – 1.85) | 0.743 |
| Group [CAI] × Timer [with] | 0.54 (0.31 – 0.97) | 0.038 | 0.63 (0.33 – 1.18) | 0.150 | 0.63 (0.32 – 1.22) | 0.169 |
| Observations | 1636 | | 1636 | | 1636 | |
| R2 Tjur | 0.096 | | 0.073 / 0.071 | | 0.084 / 0.081 | |

## C2. Regressions moderated by education

**Logistic regression of intervention moderated by education.** Odds ratios correspond to a logistic regression of the intervention group moderated by education (an ordered predictor: Low < Middle < High < Very High) and controlled for age, gender, attentiveness, and the timing condition (Timer). Models are fitted using unweighted data, raked and trimmed weights, and post-stratification weights.

| Predictors | Unweighted Odds Ratios | p | Raked and Trimmed Odds Ratios | p | Post-stratification Odds Ratios | p |
|---|---|---|---|---|---|---|
| (Intercept) | 0.81 (0.42 – 1.54) | 0.509 | 1.05 (0.52 – 2.11) | 0.900 | 0.84 (0.40 – 1.74) | 0.631 |
| Group [Search] | 2.18 (1.53 – 3.12) | **<0.001** | 2.08 (1.44 – 3.00) | **<0.001** | 2.22 (1.52 – 3.24) | **<0.001** |
| Group [CAI] | 6.31 (4.61 – 8.77) | **<0.001** | 5.68 (4.07 – 7.93) | **<0.001** | 6.48 (4.60 – 9.12) | **<0.001** |
| Education [linear] | 1.20 (0.69 – 2.15) | 0.523 | 1.07 (0.60 – 1.92) | 0.818 | 1.15 (0.64 – 2.08) | 0.639 |
| Education [quadratic] | 1.07 (0.62 – 1.83) | 0.797 | 1.05 (0.60 – 1.84) | 0.852 | 1.03 (0.58 – 1.82) | 0.920 |
| Education [cubic] | 1.09 (0.66 – 1.86) | 0.731 | 1.13 (0.66 – 1.93) | 0.651 | 1.11 (0.64 – 1.92) | 0.703 |
| Age | 1.45 (1.29 – 1.64) | **<0.001** | 1.43 (1.25 – 1.64) | **<0.001** | 1.50 (1.31 – 1.71) | **<0.001** |
| Gender [Women] | 1.42 (1.13 – 1.78) | **0.003** | 1.38 (1.07 – 1.78) | **0.013** | 1.52 (1.18 – 1.97) | **0.001** |
| Attentiveness | 0.19 (0.10 – 0.36) | **<0.001** | 0.15 (0.07 – 0.30) | **<0.001** | 0.16 (0.07 – 0.33) | **<0.001** |
| Timer [with] | 0.92 (0.73 – 1.16) | 0.465 | 1.03 (0.79 – 1.33) | 0.845 | 0.99 (0.76 – 1.29) | 0.939 |
| Group [Search] × Education [linear] | 0.30 (0.14 – 0.64) | **0.002** | 0.36 (0.16 – 0.80) | **0.012** | 0.38 (0.17 – 0.85) | **0.019** |
| Group [CAI] × Education [linear] | 0.28 (0.14 – 0.56) | **<0.001** | 0.34 (0.17 – 0.68) | **0.002** | 0.35 (0.17 – 0.71) | **0.004** |
| Group [Search] × Education [quadratic] | 0.48 (0.23 – 0.98) | **0.044** | 0.50 (0.24 – 1.03) | 0.062 | 0.46 (0.21 – 0.97) | **0.042** |
| Group [CAI] × Education [quadratic] | 0.91 (0.48 – 1.73) | 0.770 | 0.87 (0.45 – 1.68) | 0.682 | 0.87 (0.44 – 1.72) | 0.695 |
| Group [Search] × Education [cubic] | 0.82 (0.42 – 1.54) | 0.534 | 0.83 (0.43 – 1.61) | 0.575 | 0.86 (0.43 – 1.73) | 0.676 |
| Group [CAI] × Education [cubic] | 0.80 (0.44 – 1.44) | 0.465 | 0.78 (0.42 – 1.44) | 0.428 | 0.82 (0.44 – 1.54) | 0.545 |
| Observations | 1636 | | 1636 | | 1636 | |
| $R^2$ Tjur | 0.186 | | 0.138 / 0.130 | | | |

## C3. Regressions moderated by education (CSA as reference group and moderation of quadratic education term for search group)

**Logistic regression of intervention moderated by education with CSA as reference group and moderation of quadratic education term for search group.** Log-Odds corresponding to a logistic regression of the intervention group moderated by education (an ordered predictor: Low < Middle < High < Very High) and controlled for age, gender, attentiveness, and the timing condition (Timer). CSA group is used as reference. Aside the lineaized education (as moderator of the groups, we keep the moderation of the quadratic education term on Group search. Models are fitted using unweighted data, raked and trimmed weights, and post-stratification weights.

|  | Unweighted | | Raked and Trimmed | | Post-stratification | |
|---|---|---|---|---|---|---|
| Predictors | Log-Odds | p | Log-Odds | p | Log-Odds | p |
| (Intercept) | -0.19 (-0.83 – 0.46) | 0.570 | 0.04 (-0.68 – 0.75) | 0.920 | -0.18 (-0.93 – 0.57) | 0.637 |
| Group [Search] | 0.67 (0.33 – 1.01) | **<0.001** | 0.68 (0.31 – 1.04) | **<0.001** | 0.75 (0.36 – 1.14) | **<0.001** |
| Group [CAI] | 1.68 (1.39 – 1.98) | **<0.001** | 1.64 (1.31 – 1.97) | **<0.001** | 1.77 (1.43 – 2.11) | **<0.001** |
| EduL | 3.66 (-6.06 – 13.45) | 0.460 | 0.63 (-10.09 – 11.34) | 0.909 | 1.76 (-9.23 – 12.76) | 0.753 |
| Age | 0.37 (0.26 – 0.49) | **<0.001** | 0.36 (0.22 – 0.49) | **<0.001** | 0.40 (0.27 – 0.53) | **<0.001** |
| Gender [Women] | 0.34 (0.12 – 0.57) | **0.003** | 0.32 (0.07 – 0.57) | **0.014** | 0.42 (0.16 – 0.67) | **0.001** |
| Attentiveness | -1.66 (-2.31 – -1.03) | **<0.001** | -1.91 (-2.63 – -1.19) | **<0.001** | -1.84 (-2.60 – -1.09) | **<0.001** |
| Timer [with] | -0.09 (-0.32 – 0.14) | 0.454 | 0.03 (-0.24 – 0.29) | 0.849 | -0.01 (-0.27 – 0.25) | 0.935 |
| Group [Search] × EduL | -26.50 (-40.46 – -12.82) | **<0.001** | -22.21 (-36.84 – -7.58) | **0.003** | -22.15 (-37.31 – -6.99) | **0.004** |
| Group [CAI] × EduL | -23.11 (-34.83 – -11.46) | **<0.001** | -19.10 (-32.00 – -6.20) | **0.004** | -18.92 (-32.24 – -5.61) | **0.005** |
| IsSearch [NotSearch] × EduQ | -0.33 (-5.75 – 5.14) | 0.905 | -0.72 (-6.41 – 4.97) | 0.804 | -1.43 (-7.17 – 4.31) | 0.624 |
| IsSearch [Search] × EduQ | -12.50 (-21.47 – -3.78) | **0.005** | -11.95 (-20.95 – -2.96) | **0.009** | -14.13 (-23.49 – -4.78) | **0.003** |
| Observations | 1636 | | 1636 | | 1636 | |
| R² Tjur | 0.186 | | 0.137 / 0.132 | | 0.149 / 0.143 | |

## C4. Regression of intervention moderated by education, age and gender

**Logistic regression of intervention moderated by education, age and gender.** Odds ratios correspond to a logistic regression of the intervention group moderated by education (an ordered predictor: Low < Middle < High < Very High), age and gender and controlled for attentiveness, and the timing condition (Timer). Aside the lineaized education (as moderator of the groups, we keep the moderation of the quadratic education term on Group search. Models are fitted using unweighted data, raked and trimmed weights, and post-stratification weights.

| Predictors | Unweighted Log-Odds | p | Raked and Trimmed Log-Odds | p | Post-stratification Log-Odds | p |
|---|---|---|---|---|---|---|
| (Intercept) | 0.05 (-0.62 – 0.74) | 0.877 | 0.08 (-0.67 – 0.83) | 0.829 | -0.03 (-0.82 – 0.75) | 0.930 |
| Group [Search] | 0.67 (0.22 – 1.12) | **0.004** | 0.83 (0.32 – 1.35) | **0.001** | 0.83 (0.29 – 1.37) | **0.003** |
| Group [CAI] | 1.29 (0.91 – 1.70) | **<0.001** | 1.36 (0.91 – 1.81) | **<0.001** | 1.46 (0.99 – 1.93) | **<0.001** |
| EduL | -1.21 (-11.20 – 8.78) | 0.812 | -4.14 (-14.97 – 6.68) | 0.453 | -0.91 (-12.20 – 10.37) | 0.874 |
| Age | -0.18 (-0.46 – 0.09) | 0.189 | -0.18 (-0.47 – 0.12) | 0.243 | -0.07 (-0.36 – 0.23) | 0.664 |
| Gender [Women] | -0.11 (-0.63 – 0.40) | 0.669 | 0.04 (-0.52 – 0.60) | 0.886 | 0.06 (-0.51 – 0.64) | 0.824 |
| Attentiveness | -1.69 (-2.34 – -1.06) | **<0.001** | -1.88 (-2.57 – -1.19) | **<0.001** | -1.83 (-2.56 – -1.11) | **<0.001** |
| Timer [with] | -0.10 (-0.33 – 0.14) | 0.425 | 0.00 (-0.26 – 0.27) | 0.974 | -0.03 (-0.30 – 0.23) | 0.814 |
| Group [Search] × EduL | -21.75 (-35.96 – -7.79) | **0.002** | -16.96 (-31.79 – -2.12) | **0.025** | -19.25 (-34.57 – -3.94) | **0.014** |
| Group [CAI] × EduL | -16.92 (-29.03 – -4.83) | **0.006** | -13.27 (-26.72 – 0.19) | 0.053 | -16.03 (-29.96 – -2.11) | **0.024** |
| IsSearch [NotSearch] × EduQ | -0.30 (-5.81 – 5.24) | 0.914 | -0.51 (-6.29 – 5.28) | 0.864 | -1.34 (-7.22 – 4.53) | 0.654 |
| IsSearch [Search] × EduQ | -12.77 (-21.78 – -4.01) | **0.005** | -12.06 (-21.14 – -2.97) | **0.009** | -14.50 (-23.82 – -5.18) | **0.002** |
| Group [Search] × Age | 0.56 (0.21 – 0.92) | **0.002** | 0.62 (0.24 – 1.00) | **0.001** | 0.49 (0.10 – 0.89) | **0.013** |
| Group [CAI] × Age | 0.73 (0.42 – 1.05) | **<0.001** | 0.76 (0.41 – 1.10) | **<0.001** | 0.62 (0.28 – 0.96) | **<0.001** |
| Group [Search] × Gender [Women] | 0.05 (-0.62 – 0.73) | 0.882 | -0.17 (-0.89 – 0.55) | 0.646 | -0.10 (-0.85 – 0.65) | 0.797 |
| Group [CAI] × Gender [Women] | 0.81 (0.21 – 1.42) | **0.009** | 0.78 (0.13 – 1.44) | **0.019** | 0.72 (0.05 – 1.39) | **0.035** |
| Observations | 1636 | | 1636 | | 1636 | |
| $R^2$ Tjur | 0.201 | | 0.155 / 0.147 | | 0.161 / 0.153 | |

# D. Structural Equation Models

## D1. UTAUT-2 models with different versions of weights (CAI group)

**Selected structural equation models for the UTAUT-2 with different weights.** The columns show fits for the UTAUT-2 models corresponding to the weights used (first row) and whether the direct effects (D), and interaction terms (I) are present. The fit measures diagnosis for the models with interactions (D+I) indicates an acceptable fit according to cut-off values: CFI > .9, TLI > .9, RMSEA < .08, SRMR < .08, and Rel. $X^2$ < 3. Only the CAI group is considered for the avoidance (i.e., ~Avoidance section of the table).

|  | Unweighted | | Raked & Trimmed | | Post-stratification | |
|---|---|---|---|---|---|---|
|  | D Only | D + I | D Only | D + I | D Only | D + I |
| *~ BI* | | | | | | |
| *PerformanceExpectancy* | 0.31*** | 0.32*** | 0.32*** | 0.33*** | 0.35*** | 0.36*** |
| *EffortExpectancy* | -0.00 | -0.00 | -0.00 | -0.00 | -0.00 | -0.01 |
| *SocialInfluence* | 0.27*** | 0.27*** | 0.27*** | 0.27*** | 0.28*** | 0.28*** |
| *FacilCnd* | 0.04 | 0.04 | 0.05 | 0.05 | 0.05 | 0.05 |
| *HedMotiv* | 0.16*** | 0.17*** | 0.15*** | 0.16*** | 0.13*** | 0.14*** |
| *Habit* | 0.33*** | 0.33*** | 0.32*** | 0.32*** | 0.32*** | 0.32*** |
| *Age* | -0.06* | -0.06* | -0.08** | -0.08** | -0.08** | -0.08** |
| *Gender* | -0.00 | -0.00 | -0.00 | -0.00 | 0.00 | 0.00 |
| *Education* | 0.01 | 0.01 | 0.01 | 0.01 | 0.01 | 0.01 |
| *Experience* | 0.22*** | 0.22*** | 0.22*** | 0.22*** | 0.21*** | 0.21*** |
| *FacilCnd_x_Age* |  | 0.03 |  | 0.02 |  | 0.02 |
| *FacilCnd_x_Gendr* |  | -0.05 |  | -0.05 |  | -0.04 |
| *FacilCnd_x_Exp* |  | 0.01 |  | 0.01 |  | 0.00 |
| *FacilCnd_x_Edu* |  | -0.02 |  | -0.04 |  | -0.04 |
| *HedMotiv_x_Age* |  | -0.03 |  | -0.03 |  | -0.03 |
| *HedMotiv_x_Gendr* |  | 0.02 |  | 0.02 |  | 0.03 |
| *HedMotiv_x_Exp* |  | 0.02 |  | 0.01 |  | 0.02 |
| *HedMotiv_x_Edu* |  | 0.06 |  | 0.10· |  | 0.11· |
| *Habit_x_Age* |  | 0.01 |  | 0.02 |  | 0.01 |
| *Habit_x_Gendr* |  | 0.02 |  | 0.01 |  | -0.01 |
| *Habit_x_Exp* |  | -0.03 |  | -0.01 |  | -0.02 |
| *Habit_x_Edu* |  | -0.04 |  | -0.06· |  | -0.07· |
| *Attentiveness* | 0.07* | 0.07* | 0.07* | 0.07* | 0.07· | 0.07· |
| *~ Avoidance* | | | | | | |
| *BI* | -0.10 | -0.08 | -0.10 | -0.08 | -0.14 | -0.11 |
| *Age* | 0.29*** | 0.29*** | 0.32*** | 0.32*** | 0.30*** | 0.30*** |
| *Gender* | 0.18*** | 0.18*** | 0.22*** | 0.22*** | 0.22*** | 0.22*** |
| ***Education*** | ***-0.24**** | ***-0.24**** | ***-0.23**** | ***-0.23**** | ***-0.22**** | ***-0.22**** |
| *Experience* | -0.19** | -0.19** | -0.19** | -0.20** | -0.17* | -0.17* |
| *FacilCnd* | -0.19** | -0.21*** | -0.17** | -0.19** | -0.16* | -0.18** |
| *Habit* | -0.08 | -0.09 | -0.11 | -0.11 | -0.10 | -0.11 |
| *Habit_x_Age* |  | -0.01 |  | -0.04 |  | -0.02 |
| *Habit_x_Gendr* |  | -0.03 |  | -0.07 |  | -0.08 |
| *Habit_x_Exp* |  | -0.04 |  | -0.10 |  | -0.09 |
| *Habit_x_Edu* |  | -0.10 |  | -0.09 |  | -0.07 |
| *Attentiveness* | -0.14** | -0.14** | -0.14* | -0.14* | -0.15* | -0.16** |
| *Timer* | -0.19· | -0.19· | -0.17 | -0.17 | -0.17 | -0.17 |
| *$R^2$ (~ BI)* | 0.78 | 0.79 | 0.78 | 0.79 | 0.78 | 0.79 |
| *$R^2$ (~ Avoidance)* | 0.34 | 0.35 | 0.36 | 0.37 | 0.34 | 0.35 |
| *CFI* | 0.77 | 0.92 | 0.78 | 0.92 | 0.78 | 0.93 |
| *TLI* | 0.89 | 0.93 | 0.89 | 0.93 | 0.89 | 0.94 |
| *RMSEA* | 0.13 | 0.06 | 0.12 | 0.05 | 0.11 | 0.05 |
| *SRMR* | 0.02 | 0.02 | 0.03 | 0.02 | 0.03 | 0.02 |
| *Rel. $X^2$ ($X^2/df$)* | 9.81 | 2.96 | 8.3 | 2.61 | 7.74 | 2.38 |

## D2. UTAUT-2 models alternating education and experience (CAS group)

**Selected structural equation models for the UTAUT-2 with combinations of education and experience** The columns show fits for the UTAUT-2 models corresponding to variations of the models alternating the presence of experience and education (first row) and whether only direct effects are included (D only) or also interaction terms (D + I) (second row). All models used trimmed and raked weights. The fit measures diagnosis for the models with interactions (D+I) indicates an acceptable fit according to cut-off values: CFI > .9, TLI > .9, RMSEA < .08, SRMR < .08, and Rel. $X^2$ < 3. Only the CAI group is considered for the avoidance (i.e., ~Avoidance section of the table).

|  | Base | | Experience | | Education | | Experience & Education | |
|---|---|---|---|---|---|---|---|---|
|  | D Only | D + I | D Only | D + I | D Only | D + I | D Only | D + I |
| **~ BI** | | | | | | | | |
| PerformanceExpectancy | 0.32*** | 0.33*** | 0.32*** | 0.33*** | 0.32*** | 0.33*** | 0.32*** | 0.33*** |
| EffortExpectancy | -0.00 | -0.00 | -0.00 | -0.00 | -0.00 | -0.00 | -0.00 | -0.00 |
| SocialInfluence | 0.27*** | 0.27*** | 0.27*** | 0.27*** | 0.27*** | 0.27*** | 0.27*** | 0.27*** |
| FacilCnd | 0.05 | 0.04 | 0.05 | 0.04 | 0.05 | 0.04 | 0.05 | 0.05 |
| HedMotiv | 0.15*** | 0.16*** | 0.15*** | 0.16*** | 0.15*** | 0.16*** | 0.15*** | 0.16*** |
| Habit | 0.32*** | 0.32*** | 0.32*** | 0.32*** | 0.32*** | 0.32*** | 0.32*** | 0.32*** |
| Age | -0.08** | -0.08** | -0.08** | -0.08** | -0.08** | -0.08** | -0.08** | -0.08** |
| Gender | -0.00 | -0.00 | -0.00 | -0.00 | -0.00 | -0.00 | -0.00 | -0.00 |
| Education | 0.02 | 0.02 | 0.02 | 0.02 | 0.01 | 0.01 | 0.01 | 0.01 |
| Experience | 0.23*** | 0.23*** | 0.22*** | 0.22*** | 0.23*** | 0.23*** | 0.22*** | 0.22*** |
| FacilCnd_x_Age |  | 0.03 |  | 0.02 |  | 0.03 |  | 0.02 |
| FacilCnd_x_Gendr |  | -0.04 |  | -0.05 |  | -0.04 |  | -0.05 |
| FacilCnd_x_Exp |  | 0.01 |  | 0.01 |  | 0.01 |  | 0.01 |
| FacilCnd_x_Edu |  | -0.04 |  | -0.04 |  | -0.04 |  | -0.04 |
| HedMotiv_x_Age |  | -0.03 |  | -0.03 |  | -0.03 |  | -0.03 |
| HedMotiv_x_Gendr |  | 0.02 |  | 0.02 |  | 0.02 |  | 0.02 |
| HedMotiv_x_Exp |  | 0.01 |  | 0.01 |  | 0.01 |  | 0.01 |
| HedMotiv_x_Edu |  | 0.10· |  | 0.10· |  | 0.10· |  | 0.10· |
| Habit_x_Age |  | 0.02 |  | 0.02 |  | 0.01 |  | 0.02 |
| Habit_x_Gendr |  | 0.01 |  | 0.01 |  | 0.01 |  | 0.01 |
| Habit_x_Exp |  | -0.01 |  | -0.01 |  | -0.01 |  | -0.01 |
| Habit_x_Edu |  | -0.06· |  | -0.06· |  | -0.07· |  | -0.06· |
| Attentiveness | 0.07* | 0.07* | 0.07* | 0.07* | 0.07* | 0.07* | 0.07* | 0.07* |
| **~ Avoidance** | | | | | | | | |
| BI | -0.16· | -0.18· | -0.11 | -0.10 | -0.15 | -0.15 | -0.10 | -0.08 |
| Age | 0.32*** | 0.31*** | 0.32*** | 0.32*** | 0.32*** | 0.32*** | 0.32*** | 0.32*** |
| Gender | 0.22*** | 0.22*** | 0.22*** | 0.22*** | 0.22*** | 0.22*** | 0.22*** | 0.22*** |
| **Education** |  |  |  |  | **-0.23*** | -0.23*** | -0.23*** | -0.23*** |
| Experience |  |  | -0.19** | -0.19** |  |  | -0.19** | -0.20** |
| FacilCnd | -0.14* | -0.12· | -0.17* | -0.17* | -0.15* | -0.15* | -0.17** | -0.19** |
| Habit | -0.09 | -0.08 | -0.10 | -0.11 | -0.09 | -0.09 | -0.11 | -0.11 |
| Habit_x_Age |  | 0.04 |  | -0.00 |  | -0.02 |  | -0.04 |
| Habit_x_Gendr |  | -0.06 |  | -0.08 |  | -0.05 |  | -0.07 |
| Habit_x_Exp |  |  |  | -0.12· |  |  |  | -0.10 |
| Habit_x_Edu |  |  |  |  |  | -0.10 |  | -0.09 |
| Attentiveness | -0.14* | -0.14* | -0.14* | -0.14* | -0.14* | -0.14* | -0.14* | -0.14* |
| Timer | -0.17 | -0.17 | -0.17 | -0.17 | -0.17 | -0.17 | -0.17 | -0.17 |
| $R^2$ (~ BI) | 0.78 | 0.79 | 0.78 | 0.79 | 0.78 | 0.79 | 0.78 | 0.79 |
| $R^2$ (~ Avoidance) | 0.24 | 0.24 | 0.3 | 0.31 | 0.3 | 0.31 | 0.36 | 0.37 |
| CFI | 0.77 | 0.92 | 0.77 | 0.92 | 0.77 | 0.92 | 0.78 | 0.92 |
| TLI | 0.89 | 0.93 | 0.89 | 0.93 | 0.89 | 0.93 | 0.89 | 0.93 |
| RMSEA | 0.12 | 0.05 | 0.12 | 0.05 | 0.12 | 0.05 | 0.12 | 0.05 |
| SRMR | 0.03 | 0.03 | 0.03 | 0.02 | 0.03 | 0.02 | 0.03 | 0.02 |
| Rel. $X^2$ ($X^2/df$) | 8.22 | 2.62 | 8.27 | 2.62 | 8.26 | 2.61 | 8.3 | 2.61 |

## D3. UTAUT-2 models alternating education and experience (CAS group) and with or without age and gender

**Selected structural equation models for the UTAUT-2 with combinations of education and experience and with or without age and gender.** The columns show fits for the UTAUT-2 models corresponding to variations of the models alternating the presence of age and gender (first row) and of experience and education (second row). All models used trimmed and raked weights. Fit diagnosis measures for all models indicate an acceptable fit according to cut-off values: CFI > .9, TLI > .9, RMSEA < .08, SRMR < .08, and Rel. X² < 3. Only the CAI group is considered for the avoidance (i.e., ~Avoidance section of the table).

|  | Just BI | | | | Age & Gender | | | |
|---|---|---|---|---|---|---|---|---|
|  | BI Only | Experience | Education | Experience & Education | BI Only | Experience | Education | Experience & Education |
| *~ BI* | | | | | | | | |
| *PerformanceExpectancy* | 0.33*** | 0.33*** | 0.33*** | 0.33*** | 0.33*** | 0.33*** | 0.33*** | 0.33*** |
| *EffortExpectancy* | -0.01 | -0.01 | -0.01 | -0.01 | -0.01 | -0.01 | -0.01 | -0.01 |
| *SocialInfluence* | 0.26*** | 0.26*** | 0.26*** | 0.26*** | 0.26*** | 0.26*** | 0.26*** | 0.26*** |
| *FacilCnd* | 0.06 | 0.06 | 0.06 | 0.06 | 0.06 | 0.06 | 0.06 | 0.06 |
| *HedMotiv* | 0.15*** | 0.15*** | 0.15*** | 0.15*** | 0.15*** | 0.15*** | 0.15*** | 0.15*** |
| *Habit* | 0.32*** | 0.32*** | 0.32*** | 0.32*** | 0.32*** | 0.32*** | 0.32*** | 0.32*** |
| *Age* | -0.11*** | -0.11*** | -0.11*** | -0.11*** | -0.08** | -0.08** | -0.08** | -0.08** |
| *Gender* | -0.02 | -0.02 | -0.02 | -0.02 | -0.00 | -0.00 | -0.00 | -0.00 |
| *Education* | 0.03 | 0.03 | 0.01 | 0.01 | 0.03 | 0.03 | 0.01 | 0.01 |
| *Experience* | 0.23*** | 0.22*** | 0.23*** | 0.22*** | 0.23*** | 0.22*** | 0.23*** | 0.22*** |
| *FacilCnd_x_Age* | 0.04 | 0.03 | 0.04 | 0.03 | 0.03 | 0.03 | 0.03 | 0.03 |
| *FacilCnd_x_Gendr* | -0.04 | -0.04 | -0.04 | -0.04 | -0.04 | -0.04 | -0.04 | -0.04 |
| *FacilCnd_x_Exp* | 0.02 | 0.02 | 0.02 | 0.02 | 0.02 | 0.02 | 0.02 | 0.02 |
| *FacilCnd_x_Edu* | -0.04 | -0.04 | -0.04 | -0.04 | -0.04 | -0.04 | -0.04 | -0.04 |
| *HedMotiv_x_Age* | -0.03 | -0.03 | -0.03 | -0.03 | -0.03 | -0.03 | -0.03 | -0.03 |
| *HedMotiv_x_Gendr* | 0.02 | 0.02 | 0.02 | 0.02 | 0.02 | 0.02 | 0.02 | 0.02 |
| *HedMotiv_x_Exp* | 0.01 | 0.01 | 0.01 | 0.01 | 0.01 | 0.01 | 0.01 | 0.01 |
| *HedMotiv_x_Edu* | 0.10· | 0.10· | 0.10· | 0.10· | 0.10· | 0.10· | 0.10· | 0.10· |
| *Habit_x_Age* | 0.01 | 0.01 | 0.01 | 0.01 | 0.01 | 0.01 | 0.01 | 0.01 |
| *Habit_x_Gendr* | 0.01 | 0.01 | 0.01 | 0.01 | 0.01 | 0.01 | 0.01 | 0.01 |
| *Habit_x_Exp* | -0.01 | -0.01 | -0.01 | -0.01 | -0.01 | -0.01 | -0.01 | -0.01 |
| *Habit_x_Edu* | -0.06 | -0.06· | -0.06 | -0.06· | -0.06· | -0.06· | -0.06· | -0.06· |
| *Attentiveness* | 0.07* | 0.07* | 0.07* | 0.07* | 0.07* | 0.07* | 0.07* | 0.07* |
| *~ Avoidance* | | | | | | | | |
| *BI* | -0.36*** | -0.35*** | -0.36*** | -0.34*** | -0.34*** | -0.32*** | -0.33*** | -0.32*** |
| *Age* | | | | | 0.30*** | 0.30*** | 0.30*** | 0.30*** |
| *Gender* | | | | | 0.22*** | 0.22*** | 0.22*** | 0.22*** |
| **Education** | | | **-0.23*** | **-0.23*** | | | **-0.23*** | **-0.23*** |
| *Experience* | | -0.14· | | -0.14· | | -0.15* | | -0.15* |
| *Attentiveness* | -0.12* | -0.12* | -0.12* | -0.12* | -0.13* | -0.13* | -0.13* | -0.13* |
| *Timer* | -0.17 | -0.17 | -0.17 | -0.17 | -0.17 | -0.17 | -0.17 | -0.17 |
| *R² (~ BI)* | 0.8 | 0.8 | 0.8 | 0.8 | 0.8 | 0.8 | 0.8 | 0.79 |
| *R² (~ Avoidance)* | 0.13 | 0.18 | 0.19 | 0.24 | 0.26 | 0.31 | 0.32 | 0.37 |
| *CFI* | 0.92 | 0.92 | 0.92 | 0.92 | 0.92 | 0.92 | 0.92 | 0.92 |
| *TLI* | 0.93 | 0.93 | 0.93 | 0.93 | 0.93 | 0.93 | 0.93 | 0.93 |
| *RMSEA* | 0.06 | 0.06 | 0.06 | 0.06 | 0.05 | 0.05 | 0.05 | 0.05 |
| *SRMR* | 0.03 | 0.03 | 0.03 | 0.03 | 0.03 | 0.03 | 0.03 | 0.03 |
| *Rel. X² (X²/df)* | 2.66 | 2.66 | 2.66 | 2.66 | 2.64 | 2.64 | 2.64 | 2.64 |

## D4. Comparison: manifest indicators or latent factor loadings on single indicator items

To incorporate prior information from the original UTAUT-2 measurement model, we conducted an additional SEM analysis in which the factor loadings (λ) for social influence, habit, and behavioral intention were fixed to 0.80, 0.84, and 0.87, respectively, as reported in Venkatesh et al. (2012). Correspondingly, we set the measurement error variance to $1 - \lambda^2$. To ensure proper model identification and to prevent estimation issues (e.g., negative variance estimates), we constrained the variance of the latent constructs to 1 (Kline, 2016; Rosseel, 2012).

**Comparison of models using manifest indicators or latent factor loadings on single indicator items (CAI group).** The columns show models using combinations of UTAUT-2 variables corresponding to (a) baselines that only included demographics with or without experience, (b) models that ignore the Behavioural Intention (BI) path, and (c) the full UTAUT-2 model. For (b) and (c), they either use the manifest indicators or the latent factor loading. All models used trimmed and raked weights (second row). Only the fit measures diagnosis for the full model with the manifest indicators is acceptable according to cut-off values: CFI > .9, TLI > .9, RMSEA < .08, SRMR < .08, and Rel. $X^2$ < 3. Only the CAI group is considered for the avoidance (i.e., ~Avoidance section of the table).

|  | (a) Baselines | | | (b) Skip BI path | | (c) Full | |
|---|---|---|---|---|---|---|---|
|  | Demographics | + BI | + BI & Experience | Manifest | Factor Loadings | Observed | Factor Loadings |
| **~ BI** | | | | | | | |
| *PerformanceExpectancy* | | | | | | 0.33*** | 0.21*** |
| *EffortExpectancy* | | | | | | -0.00 | 0.10· |
| *SocialInfluence* | | | | | | 0.27*** | 0.06* |
| *FacilCnd* | | | | | | 0.05 | 0.08 |
| *HedMotiv* | | | | | | 0.16*** | 0.18*** |
| *Habit* | | | | | | 0.32*** | 0.07** |
| *Age* | | | | | | -0.08** | -0.20*** |
| *Gender* | | | | | | -0.00 | -0.09* |
| *Education* | | | | | | 0.01 | 0.06 |
| *Experience* | | | | | | 0.22*** | 0.59*** |
| *FacilCnd_x_Age* | | | | | | 0.02 | 0.02 |
| *FacilCnd_x_Gendr* | | | | | | -0.05 | -0.03 |
| *FacilCnd_x_Exp* | | | | | | 0.01 | 0.01 |
| *FacilCnd_x_Edu* | | | | | | -0.04 | -0.02 |
| *HedMotiv_x_Age* | | | | | | -0.03 | -0.03 |
| *HedMotiv_x_Gendr* | | | | | | 0.02 | 0.01 |
| *HedMotiv_x_Exp* | | | | | | 0.01 | -0.02 |
| *HedMotiv_x_Edu* | | | | | | 0.10· | 0.03 |
| *Habit_x_Age* | | | | | | 0.02 | 0.01 |
| *Habit_x_Gendr* | | | | | | 0.01 | 0.01 |
| *Habit_x_Exp* | | | | | | -0.01 | 0.04 |
| *Habit_x_Edu* | | | | | | -0.06· | -0.02 |
| *Attentiveness* | | | | | | 0.07* | 0.01 |
| **~ Avoidance** | | | | | | | |
| *BI* | | -0.22*** | -0.13· | -0.15· | -0.03 | -0.08 | -0.02 |
| *Age* | 0.35*** | 0.29*** | 0.29*** | 0.29*** | 0.31*** | 0.32*** | 0.31*** |
| *Gender* | 0.24*** | 0.21*** | 0.20*** | 0.20*** | 0.21*** | 0.22*** | 0.21*** |
| **Education** | **-0.26*** | **-0.23*** | **-0.22*** | **-0.22*** | **-0.23*** | **-0.23*** | **-0.22*** |
| *Experience* | | | -0.16* | -0.17* | -0.23*** | -0.20** | -0.21*** |
| *FacilCnd* | | | | -0.14* | -0.16* | -0.19** | -0.16* |
| *Habit* | | | | 0.04 | 0.02 | -0.11 | 0.06 |
| *Habit_x_Age* | | | | | | -0.04 | -0.07 |
| *Habit_x_Gendr* | | | | | | -0.07 | -0.06 |
| *Habit_x_Exp* | | | | | | -0.10 | -0.10 |
| *Habit_x_Edu* | | | | | | -0.09 | -0.07 |
| *Attentiveness* | -0.13* | -0.14* | -0.15* | -0.14* | -0.15** | -0.14* | -0.15** |
| *Timer* | -0.18 | -0.17 | -0.17 | -0.18 | -0.18 | -0.17 | -0.18 |
| $R^2$ (~ BI) | | | | | | 0.79 | 0.43 |
| $R^2$ (~ Avoidance) | 0.23 | 0.26 | 0.27 | 0.28 | 0.28 | 0.37 | 0.3 |
| CFI | | 1 | 1 | 0 | 0 | 0.92 | 0.87 |
| TLI | | 1 | 1 | 0.73 | 0.14 | 0.93 | 0.86 |
| RMSEA | | 0 | 0 | 0.14 | 0.23 | 0.05 | 0.08 |
| SRMR | | 0 | 0 | 0.02 | 0.01 | 0.02 | 0.03 |
| Rel. $X^2$ ($X^2$/df) | | | Inf | 12.28 | 29.67 | 2.61 | 4.22 |

## D5. Structural equation models of the Behavioural Intention (BI) path of the UTAUT-2

**Structural equation models of the Behavioural intention (BI) path of the UTAUT-2.** The columns show fits for the Behavioural Intention (BI) path corresponding to variations of the models alternating the presence of experience and education (first row) and whether only direct effects are included (D only) or also interaction terms (D + I) (second row). All models used trimmed and raked weights.

|  | Base | | Experience | | Education | | Experience & Education | |
|---:|:---:|:---:|:---:|:---:|:---:|:---:|:---:|:---:|
|  | D Only | D + I | D Only | D + I | D Only | D + I | D Only | D + I |
| *~ BI* | | | | | | | | |
| *PerformanceExpectancy* | 0.47*** | 0.47*** | 0.45*** | 0.44*** | 0.47*** | 0.46*** | 0.45*** | 0.44*** |
| *EffortExpectancy* | -0.06 | -0.06 | -0.04 | -0.04 | -0.06 | -0.06 | -0.04 | -0.04 |
| *SocialInfluence* | 0.05* | 0.06* | 0.05* | 0.06* | 0.05· | 0.06* | 0.05* | 0.06* |
| *FacilCnd* | 0.07 | 0.07 | 0.03 | 0.02 | 0.06 | 0.06 | 0.02 | 0.01 |
| *HedMotiv* | 0.24*** | 0.23*** | 0.25*** | 0.25*** | 0.24*** | 0.25*** | 0.25*** | 0.27*** |
| *Habit* | 0.21*** | 0.22*** | 0.17*** | 0.18*** | 0.21*** | 0.21*** | 0.17*** | 0.17*** |
| *Age* | -0.00 | 0.00 | 0.00 | 0.01 | 0.01 | 0.01 | 0.01 | 0.02 |
| *Gender* | -0.01 | -0.01 | -0.00 | -0.00 | -0.01 | -0.01 | -0.01 | -0.01 |
| **Education** | | | | | **0.05\*\*\*** | **0.05\*\*\*** | **0.05\*\*\*** | **0.05\*\*\*** |
| *Experience* | | | 0.10*** | 0.10*** | | | 0.10*** | 0.10*** |
| *FacilCnd_x_Age* | | 0.01 | | 0.00 | | 0.01 | | -0.00 |
| *FacilCnd_x_Gendr* | | -0.02 | | -0.03 | | -0.02 | | -0.03 |
| *FacilCnd_x_Exp* | | | | -0.02 | | | | -0.02 |
| *FacilCnd_x_Edu* | | | | | | 0.00 | | -0.00 |
| *HedMotiv_x_Age* | | -0.03 | | -0.02 | | -0.02 | | -0.01 |
| *HedMotiv_x_Gendr* | | -0.02 | | -0.01 | | -0.02 | | -0.02 |
| *HedMotiv_x_Exp* | | | | 0.00 | | | | -0.00 |
| *HedMotiv_x_Edu* | | | | | | 0.04· | | 0.05* |
| *Habit_x_Age* | | 0.03 | | 0.03 | | 0.03 | | 0.02 |
| *Habit_x_Gendr* | | 0.01 | | 0.01 | | 0.01 | | 0.01 |
| *Habit_x_Exp* | | | | -0.01 | | | | -0.01 |
| *Habit_x_Edu* | | | | | | -0.04· | | -0.04· |
| *Attentiveness* | -0.00 | -0.01 | -0.00 | 0.00 | -0.01 | -0.01 | -0.00 | -0.00 |
| *R² (~ BI)* | 0.71 | 0.72 | 0.71 | 0.71 | 0.71 | 0.72 | 0.71 | 0.72 |
| *CFI* | 0.92 | 0.92 | 0.91 | 0.91 | 0.91 | 0.92 | 0.91 | 0.91 |
| *TLI* | 0.88 | 0.89 | 0.88 | 0.88 | 0.88 | 0.89 | 0.88 | 0.88 |
| *RMSEA* | 0.13 | 0.09 | 0.12 | 0.08 | 0.12 | 0.08 | 0.12 | 0.08 |
| *SRMR* | 0.25 | 0.15 | 0.27 | 0.14 | 0.24 | 0.13 | 0.26 | 0.13 |
| *Rel. X² (X²/df)* | 27.8 | 14.19 | 26.29 | 12.33 | 25.47 | 11.61 | 24.23 | 10.44 |

# E. Cronbach Alpha for multiple-items indicators of UTAUT-2 and affective variables.

**Cronbach Alpha for multiple-items indicators of UTAUT-2 and affective variables.** The column shows respectively the name of the variable, followed by the cronbach alpha and confidence intervals calculated using the Feldt method and bootstrapped.

|  | Cronbach Alpha | Feldt Lower | Feldt Upper | Bootstrapped Lower | Bootstrapped Upper |
|---|---|---|---|---|---|
| *Performance Expectancy (n=2)* | 0.916 | 0.907 | 0.923 | 0.902 | 0.927 |
| *Effort Expectancy (n=2)* | 0.921 | 0.913 | 0.929 | 0.907 | 0.933 |
| *Facilitation Conditions (n=2)* | 0.918 | 0.909 | 0.925 | 0.904 | 0.930 |
| *Hedonistic Motivation (n=2)* | 0.965 | 0.961 | 0.968 | 0.959 | 0.970 |
| *Trust (n=2)* | 0.919 | 0.911 | 0.927 | 0.907 | 0.930 |
| *Anxiety (n=3)* | 0.714 | 0.689 | 0.737 | 0.684 | 0.742 |

# F. LASSO regression of affective variables on avoidance

**LASSO and logistic regression on CAI avoidance.** Results from LASSO regression with 10-fold cross-validation and corresponding logistic regression models. LASSO was used to select variables, while logistic regression provided more interpretable coefficient estimates to assess robustness. The table includes variable selection using the optimal regularization parameter that minimizes the mean squared error ($\lambda_{min}$) and the more parsimonious model corresponding to the largest lambda value within one standard error of the minimum error ($\lambda_{1se}$). Models are fitted using unweighted data, raked and trimmed weights, and post-stratification weights.

|  | Unweighted | | | | Raked and Trimmed | | | | Poststratification | | | |
|---|---|---|---|---|---|---|---|---|---|---|---|---|
|  | $\lambda_{min}$ | | $\lambda_{1se}$ | | $\lambda_{min}$ | | $\lambda_{1se}$ | | $\lambda_{min}$ | | $\lambda_{1se}$ | |
|  | LASSO | GLM | LASSO | GLM | LASSO | GLM | LASSO | GLM | LASSO | GLM | LASSO | GLM |
|---|---|---|---|---|---|---|---|---|---|---|---|---|
| *Gender* | 0.05 | 0.212*** | 0.02 | 0.222*** | 0.06 | 0.173** | 0.03 | 0.190** | 0.06 | 0.160* | 0.03 | 0.174** |
| *Age* | 0.09 | 0.287*** | 0.06 | 0.277*** | 0.09 | 0.241*** | 0.07 | 0.240*** | 0.09 | 0.257*** | 0.07 | 0.256*** |
| *Education* | -0.08 | -0.216*** | -0.06 | -0.228*** | -0.07 | -0.181** | -0.05 | -0.182** | -0.07 | -0.173* | -0.05 | -0.172* |
| *Experience* | -0.07 | -0.216** | -0.05 | -0.208** | -0.07 | -0.226** | -0.06 | -0.223** | -0.07 | -0.221** | -0.05 | -0.220** |
| *Trust* |  |  |  |  |  |  |  |  |  |  |  |  |
| *Anxiety* |  |  |  |  |  |  |  |  |  |  |  |  |
| *Negativity* | 0.01 | 0.031 |  |  | 0.01 | 0.034 | 0.00 | 0.030 | 0.01 | 0.015 |  |  |
| *Optimism* | -0.03 | -0.082 | -0.02 | -0.109· | -0.02 | -0.042 | -0.01 | -0.060 | -0.02 | -0.066 | -0.02 | -0.076 |
| *Pessimism* | 0.01 | 0.030 |  |  | 0.02 | 0.054 | 0.01 | 0.036 | 0.03 | 0.078 | 0.01 | 0.078 |
| *Profit* |  |  |  |  |  |  |  |  |  |  |  |  |
| *Disadvantage* |  |  |  |  | 0.01 | -0.033 |  |  | 0.01 | -0.015 |  |  |
| *Attentiveness* | -0.04 | -0.092 |  |  | -0.04 | -0.061 | -0.00 | -0.069 | -0.04 | -0.113· | -0.01 | -0.121· |
| *Timer* | -0.03 | -0.162** |  |  | -0.02 | -0.163* |  |  | -0.02 | -0.139* |  |  |

# G. Survey items

**Items of the study.** The columns indicate the name of the variable in the text (first), the name of the column in the data (second), the English version of the item in the survey (third), and a description of the values that the variable can hold (fourth).

| Variable | Column | Item | Value |
|---|---|---|---|
| *diversity* | bot_diversity | Which chatbots have you used? Select all that apply: ChatGPT, BingChat, Bard, Perplexity.ai, You.com, Others (blanks space) | Sum of selected: [0 to 6] |
| *applications* | bot_experiences | Please indicate how you have used chatbots in the last six months. Select all that apply. (1) Content creation and writing (creative, academic, journaling), (2) Problem solving (brainstorming, math), (3) Entertainment (chatting, roleplaying, memes), (4) Language and communication (translations, practicing foreign language), (5) Practical assistance (event planning, self-therapy), (6) Programming (7) Information search, concept learning (encyclopedic, historical), (8) Information search (news, current events), (9) Other purposes, please list all that apply: (give at least 3 open lines) | Sum of selected: [0-9] |
| *experience* | bot_experience | I have not used chatbots [in the previous item] | binary |
| *behavioural intention* | utaut_BI | I intend to use chatbots in the future. | 7 point likert scale (disagree to agree) |
| *performance expectancy* | utaut_perf1 | I believe chatbots are helpful in everyday life. | |
| *performance expectancy* | utaut_perf2 | Using chatbots helps people get tasks and projects done faster. | |
| *effort expectancy* | utaut_effort1 | It is (would be) easy for me to become skillful at using chatbots. | |
| *effort expectancy* | utaut_effort2 | I find (would find) using chatbots easy. | |
| *social influence* | utaut_social | People who are important to me think I should use chatbots. | |
| *facilitating conditions* | utaut_facilit1 | I have the knowledge necessary to use chatbots. | |
| *facilitating conditions* | utaut_facilit2 | I have the resources necessary to use chatbots. | |
| *hedonistic motivation* | utaut_hedonic1 | I think using chatbots is fun. | |
| *hedonistic motivation* | utaut_hedonic2 | I think using chatbots is enjoyable. | |
| *habit* | utaut_habit | The use of chatbots is a habit for me. | |
| *negativism* | bot_expct_neg | When you think about the use of AI in chatbots, to what extent are you concerned? | not at all, slightly, moderately, very, extremely worried |
| *optimism* | bot_expct_opt | How optimistic, if at all, are you when you think about this use of AI? | |
| *pessimism* | bot_expct_pess | How pessimistic, if at all, are you when you think about this use of AI? | |
| *perceived advantages* | bot_expct_adv | If you experience this use of AI in your environment, how much do you personally benefit from the consequences? | |
| *perceived disadvantages* | bot_expct_dis | If you experience this use of AI in your environment, how much are you personally disadvantaged by the consequences? | |
| *anxiety* | bot_anx1 | I find chatbots intimidating. | 7 point likert scale (disagree to agree) |
| *anxiety* | bot_anx2 | I am afraid that widespread use of chatbots will take jobs away from people. | |
| *anxiety* | bot_anx3 | I am afraid that if I begin to use chatbots I will become dependent upon them and lose some of my reasoning skills. | |
| *trust* | bot_trust1 | I feel secure that when I rely on the chatbot, I will get the right answers. | |
| *trust* | bot_trust2 | I am confident in the chatbot. I feel that it works well. | |